\documentclass[11pt]{article}
\usepackage{amssymb}

\def\draftversion{N}

\if \draftversion Y

\fi

\newcommand{\reseteqnum}{\setcounter{equation}{0}}

\newcommand{\plb}[3]{Phys. Lett. {\bf B#1} (#2) #3} 
\newcommand{\prl}[3]{Phys. Rev. Lett. {\bf #1} (#2) #3}
\newcommand{\prd}[3]{Phys. Rev. {\bf D#1} (#2) #3}
\newcommand{\npb}[3]{Nucl. Phys. {\bf B#1} (#2) #3}
\newcommand{\npbps}[3]{Nucl. Phys. {\bf B}(Proc. Suppl.) {\bf #1} (#2) #3}

\title{
\hfill
\parbox{3cm}{\normalsize KUNS-1560\\
{\tt  hep-lat/9902022}}\\
\vspace{0.5cm}
Low energy effective action of domain-wall fermion
and the Ginsparg-Wilson relation
\author{
Yoshio Kikukawa\thanks{e-mail address:
kikukawa@gauge.scphys.kyoto-u.ac.jp} \ 
and \ Tatsuya Noguchi\thanks{e-mail address:
noguchi@gauge.scphys.kyoto-u.ac.jp} 
\\
{\normalsize\em Department of Physics, Kyoto University 
}\\
{\normalsize\em Kyoto 606-8502, Japan}
\\
\\
\date{\normalsize February, 1999}
}
}
\begin{document}
\maketitle

\begin{abstract}
We derive the effective action of the light fermion field 
of the domain-wall fermion, which is referred as $q(x)$ by 
Furman and Shamir. The inverse of the effective Dirac operator 
turns out to be identical to the inverse of the truncated overlap 
Dirac operator, except a local contact term which would give the 
chiral symmetry breaking in the Ginsparg-Wilson relation. 
This result allows us to relate the light fermion field and the 
fermion field described by the truncated overlap Dirac operator 
and to understand the chiral property of the light fermion through 
the exact chiral symmetry based on the Ginsparg-Wilson relation.
\end{abstract}
\newpage

\section{Introduction} 
\label{sec:introduction}
\reseteqnum

Recently, our understanding of chiral symmetry on the lattice
has substantially improved. 
Lattice Dirac operators 
have been obtained \cite{overlap-D,fixed-point-D},
which are gauge covariant, 
define local actions \cite{locality-of-overlap-D} and 
satisfy the Ginsparg-Wilson relation \cite{ginsparg-wilson-rel}.
The Ginsparg-Wilson relation 
\begin{equation}
  \gamma_5 D + D \gamma_5 = a D R \gamma_5 D 
\end{equation}
implies the exact chiral symmetry of the action
under the transformation \cite{exact-chiral-symmetry}
\begin{equation}
\label{eq:chiral-transformation-of-psi}
  \delta \psi(x) = \gamma_5 \left( 1-aRD \right) \psi(x), \quad
  \delta \bar \psi(x) = \bar \psi(x) \gamma_5 .
\end{equation}

The explicit gauge covariant solution of the Ginsparg-Wilson relation 
has been derived by Neuberger \cite{overlap-D} from the overlap 
formulation of chiral determinant \cite{overlap}. 
It is defined through
the hermitian Wilson-Dirac operator $H$ with a negative mass
in a certain range
\footnote{We adopt the definition of the overlap Dirac
operator so that the normalization of the factor one half is 
included. This leads to the Ginsparg-Wilson relation with $R=2$.},
\begin{equation}
\label{eq:overlap-dirac-operator}
  D= \frac{1}{2a } \left( 1+\gamma_5\frac{H}{\sqrt{H^2}}\right) .
\end{equation}
The locality properties of the Dirac operator has been 
examined by Hern\'andes, Jansen and L\"uscher \cite{locality-of-overlap-D}.
The issue of the practical implimentation of the Dirac operator
has been studied by Neuberger \cite{practical-D-neuberger}, 
Edwards, Heller and Narayanan \cite{practical-D-narayanan-etal}, 
Chiu \cite{practical-D-chiu} 
and A.~Borici \cite{practical-D-borici}.

The domain-wall fermion \cite{domain-wall-fermion} is the basis 
of the overlap Dirac operator.
In a simplified formulation 
\cite{boundary-fermion,boundary-fermion-QCD,truncated-overlap},
the domain-wall fermion consists of $N$-flavor Wilson 
fermions with a certain flavor-mixing mass matrix.
Due to its structure of the chiral hopping and the boundary condition 
in the flavor space, a single light Dirac fermion can emerge in the 
spectrum. This light fermion can be probed most suitably
by the field variables at the boundary in the flavor space, 
which are referred as $q(x)$ and $\bar q(x)$ by Furman and Shamir.
It has been argued that the chiral symmetry of 
the light fermion is preserved up to corrections suppressed 
exponentially in the number of flavors \cite{boundary-fermion-QCD}. 
This fact has been observed 
numerically \cite{vranas-schwinger-model,blum-soni} 
and have been found useful for the numerical simulation of 
lattice QCD \cite{columbia,blum-soni-wingate,lagae-sinclair}.
The perturbative studies are found in 
\cite{aoki-taniguchi,kikukawa-neuberger-yamada}.
The authors refer the reader to \cite{blum-lat98} for 
recent review. 
In this context, the chiral transformation of this light fermion
is defined as
\begin{equation}
\label{eq:chiral-transformation-of-q}
\delta q(x) = \gamma_5 q(x), \quad
\delta \bar q(x) = \bar q(x) \gamma_5 .
\end{equation}

The goal of this paper is to understand the chiral property
of the light fermion of the domain-wall fermion 
from the point of view of the exact chiral 
symmetry based on the Ginsparg-Wilson relation.
Several authors have discussed 
the direct relation between the domain-wall fermion and the Dirac 
fermion which is described by the overlap Dirac operator.
Vranas has shown that 
in order for the subtraction of the massive modes of the domain-wall 
fermion, it is suitable to introduced the $N$-flavor 
Wilson-Dirac boson 
with the flavor-mixing mass matrix which is anti-periodic 
in the flavor space \cite{vranas-pauli-villars}.
Neuberger has shown in \cite{truncated-overlap} through the 
explicit calculation of the partition function how the overlap 
Dirac operator emerges in the limit of infinite number of flavors:
the subtracted partition function at a finite flavor $N$ can be written 
as a single determinant of the truncated overlap Dirac operator
\begin{equation}
\label{eq:truncated-overlap-dirac-operator}
D_N= 
\frac{1}{2a}\left(
1+\gamma_5 \tanh \frac{N}{2} a_5 \widetilde H \right). 
\end{equation}
Note that $\widetilde H$ here is defined through the transfer matrix of the
five-dimensional Wilson fermion with a negative mass:
\begin{equation}
\widetilde H = - \frac{1}{a_5} \ln T .
\end{equation}
In the limit of infinite flavors, this reduces to the overlap
Dirac operator Eq.~(\ref{eq:overlap-dirac-operator}), in which
the hermitian Wilson-Dirac operator $H$ is replaced by $\widetilde H$. 
Thus the domain-wall fermion with the subtraction
of the Pauli-Villars field is equivalent to the Dirac fermion 
described by the truncated overlap Dirac operator.

The relation between the light fermion field $q(x)$ and $\bar q(x)$  
of the domain-wall fermion and the fermion field 
$\psi(x)$ and $\bar \psi(x)$ described by the overlap Dirac operator
has also been suggested by several authors. 
Neuberger pointed out in \cite{truncated-overlap} 
the correspondence between the mass term
for $q(x)$ 
\begin{equation}
  m \, \bar q(x) q(x) 
\end{equation}
and the mass term for $\psi(x)$
\begin{equation}
  m \, \bar \psi(x) \left(1-\frac{a}{2}RD\right) \psi(x) .
\end{equation}
It has also been noticed by the authors of 
\cite{chiu-etal,neidermayer-lat98}
that the transformation property of the operator 
$\left(1-\frac{a}{2}RD\right) \psi(x)$ under the chiral transformation 
Eq.~(\ref{eq:chiral-transformation-of-psi}) is same as $q(x)$:
\begin{equation}
  \delta \left\{ \left(1-\frac{a}{2}RD\right) \psi(x) \right\} 
= \gamma_5 \left\{ \left(1-\frac{a}{2}RD\right) \psi(x) \right\}  .
\end{equation}
These facts suggest that there could be a correspondence as 
\begin{eqnarray}
q(x), \quad \bar q(x) 
&\Longleftrightarrow& 
\left(1-\frac{a}{2}RD\right) \psi(x) ,  \quad \bar \psi(x) 
\end{eqnarray}
at least in the limit of infinite flavors.\footnote{This 
correspondence could be different if we adopt a different transformation 
of the exact chiral symmetry of L\"uscher from 
Eq.~(\ref{eq:chiral-transformation-of-psi}).} 

In this paper, we will further examine the above correspondence
between the light fermion field of the domain-wall fermion and 
the Dirac field described by the (truncated) overlap Dirac 
operator. 
For this purpose, we derive the low energy effective action 
of the light fermion field by integrating out 
$N-1$ heavy flavors of the domain-wall fermion:
\begin{equation}
  S_N^{\rm eff} = a^4 \sum_x \bar q(x) \, D_N^{\rm eff} \, q(x). 
\end{equation}
As easily understood, this can be achieved by calculating 
the propagator of $q(x)$ and $\bar q(x)$, 
because it should be given by the inverse 
of the effective Dirac operator of these fields. 
\footnote{ In this paper, the bra-ket symbol $\langle \cdots 
\rangle$ denotes the Wick contraction of the fermion fields in it
by their propagators, not the fermionic vacuum expectation values
which must includes the weight of the fermion action.
The bra-ket symbol with the subscript $c$ denotes 
the connected contraction. 
}
\begin{equation}
  \left\langle q(x) \, \bar q(y) \right\rangle 
=  \frac{1}{a^4} { D_N^{\rm eff} }^{-1}(x,y) .
\end{equation}
It turns out that 
the propagator of the light fermion
is closely 
related to the inverse of the truncated overlap Dirac operator as follows:
\begin{eqnarray}
\frac{a}{a_5} { D_N^{\rm eff} }^{-1}+ a \delta(x,y) 
&=& 
{ D_N^{\rm \phantom{f}} }^{-1} .
\end{eqnarray}
Namely, 
the inverse of the effective Dirac operator 
gives the inverse of the truncated overlap Dirac operator up to 
a local contact term. This contact term just takes account of the 
chiral symmetry breaking in the Ginsparg-Wilson relation, 
which holds true for the overlap Dirac operator 
in the limit of the infinite flavors. 

The above relation 
allows us to relate the field variables of the light fermion, 
$q(x)$ and $\bar q(x)$, with the field variables
described by the truncated overlap Dirac operator, 
$\psi(x)$ and $\bar \psi(x)$. 
Then we can clarify the relation between the 
almost preserved chiral symmetry of the domain-wall fermion under 
the transformation Eq.~(\ref{eq:chiral-transformation-of-q}) and the 
(would-be) exact chiral symmetry of the Dirac fermion described 
by the (truncated) overlap Dirac operator under the transformation 
Eq.~(\ref{eq:chiral-transformation-of-psi}).
It is also possible to 
relate the low energy observables of the domain-wall 
fermion which are written in terms of $q(x)$ and $\bar q(x)$ 
to those written in terms of $\psi(x)$ and $\bar \psi(x)$ 
\begin{equation}
  {\cal O}_{\rm DW} [q,\bar q] = {\cal O}_N [\psi, \bar \psi; D_N ] 
\end{equation}
and to examine the chiral properties of the observables
through the (would-be) exact chiral symmetry based 
on the Ginsparg-Wilson relation.

This article is organized as follows. 
In section~\ref{sec:effective-action-of-q}, 
we will derive the effective action of the light fermion field
of the domain-wall fermion by integrating out the heavy 
$N-1$ flavors. 
The calculation of the effective action (the propagator of the light
fermion) 
is a straightforward application of that given by Neuberger 
in \cite{truncated-overlap}.
In section~\ref{sec:chiral-property-q-and-psi}, 
we will establish the relation between the light fermion field
and the fermion field described by the truncated overlap Dirac operator. 
Then we discuss the chiral properties of the light fermion 
from the point of view of the exact chiral symmetry based 
on the Ginsparg-Wilson relation.
In section~\ref{sec:axial-anomaly-q-and-psi}, 
we reexamine the axial anomaly of the domain-wall fermion 
\cite{axial-anomaly-in-domain-wall}
in view of the axial anomaly associated with the exact chiral symmetry 
\cite{index-theorem-at-finite-lattice,exact-chiral-symmetry,
lattice-chiral-jaccobian}.
In section~\ref{sec:Pauli-Villars-field}, the contribution of 
the Pauli-Villars field to the currents and the axial anomaly 
is examined.

\section{Low energy effective action of the domain-wall fermion}
\label{sec:effective-action-of-q}
\reseteqnum

We first review briefly the domain-wall fermion and its relation 
to the Dirac fermion described by the overlap Dirac operator. 
Then we evaluate the low energy effective action of the light 
fermion of the domain-wall fermion and discuss the result
in relation to the Ginsparg-Wilson relation.

\subsection{Light fermion of the domain-wall fermion}
The domain-wall fermion, in its simplified 
formulation \cite{boundary-fermion,boundary-fermion-QCD,truncated-overlap},
consists of $N$-flavor Wilson fermions\footnote{In this paper, 
we assume that the number of flavor $N$ is even.}
\begin{equation}
\label{eq:action-domain-wall-fermion}
S_{\rm DW}
=\sum_{s,t=1}^N a^4 \sum_x
\bar \psi_{s}(x) 
\left\{ 
 \gamma_\mu \frac{1}{2}\left(\nabla_\mu+\nabla_\mu^\ast\right) \delta_{st} 
+ P_L M_{st} + P_R M^\dagger_{st} 
\right\} \psi_{t}(x) 
\end{equation}
with a certain flavor-mixing mass matrix:
in the case with $N=6$, it is given by
\begin{eqnarray}
\label{eq:flavor-mixing-mass-matrix}
M_{st}&=&
\frac{1}{a_5}
\left( \begin{array}{cccccc}
              B & -1 & 0 & 0 & 0 & 0 \\
              0 & B & -1 & 0 & 0 & 0 \\
              0 & 0 & B & -1 & 0 & 0 \\
              0 & 0 & 0 & B & -1 & 0 \\
              0 & 0 & 0 & 0 & B  & -1 \\
              0 & 0 & 0 & 0 & 0 & B  
\end{array} \right),
\end{eqnarray}
and
\begin{eqnarray}
\label{eq:operator-B}
 B &=& 1 + a_5 \left( 
-\frac{a}{2} \nabla_\mu\nabla_\mu^\ast - \frac{m_0}{a} 
\right) .
\end{eqnarray}
Due to its structure of the chiral hopping and the boundary condition 
in the flavor space, a single light Dirac fermion can emerge in the 
spectrum. The exact eigenvalues and eigenvectors 
of the mass matrix for the free theory at a finite flavor $N$ has 
been given in \cite{truncated-overlap}.
This light fermion can be probed most suitably
by the field variables at the boundary in the flavor space
and is denoted as $q(x)$ and $\bar q(x)$ by Furman and Shamir:
\begin{equation}
  q(x) = \psi_{1L}(x) + \psi_{NR}(x) , \quad 
 \bar q(x) = \bar \psi_{1L}(x) + \bar \psi_{NR}(x).
\end{equation}

Following Neuberger, we may change the flavor index of the left-handed 
component by the chirally asymmetric parity transformation in the 
flavor space:
\begin{eqnarray}
  \psi^\prime_s(x) &=& \left( P_R + P_L P \right)_{st} \psi_t(x),  \\
  \bar \psi^\prime_s(x) &=& \bar \psi_t(x) \left( P_R P + P_L \right)_{ts} ,
\end{eqnarray}
where
\begin{equation}
  P_{st} = \left( \begin{array}{cccccc}
          0 & 0 & 0 & 0 & 0 & 1 \\
          0 & 0 & 0 & 0 & 1 & 0 \\
          0 & 0 & 0 & 1 & 0 & 0 \\
          0 & 0 & 1 & 0 & 0 & 0 \\
          0 & 1 & 0 & 0 & 0 & 0 \\
          1 & 0 & 0 & 0 & 0 & 0 
 \end{array} \right) \quad (N=6). \\
\end{equation}
By this transformation, the mass matrix becomes hermitian,
\begin{eqnarray}
\label{eq:hermitian-mass-matrix}
M^{\rm H}_{st}&=& M_{st} P =P M^\dagger_{st}  \nonumber\\ 
&=&
\frac{1}{a_5}
\left( \begin{array}{cccccc}
           0 & 0 & 0 & 0 & -1& B\\
           0 & 0 & 0 & -1& B & 0 \\
           0 & 0 & -1& B & 0 & 0 \\
           0 &-1 & B & 0 & 0 & 0 \\
           -1& B & 0 & 0 & 0 & 0 \\ 
           B & 0 & 0 & 0 & 0 & 0 
\end{array} \right) , \quad  (N=6). \nonumber\\
\end{eqnarray}
In this basis, it is the $N$-th flavor field that is most suitable 
to probe the light fermion:
\begin{equation}
  q(x) = \psi_N^\prime(x), \qquad \bar q(x) = \bar \psi_N^\prime(x) .
\end{equation}

The chiral transformation adopted by Shamir and Furman 
\cite{boundary-fermion-QCD} is given in this hermitian basis as follows:
\begin{equation}
\label{eq:chiral-transformation-of-DW}
\delta \psi'_s(x) 
= \left( \Gamma_5 \right)_{st} \psi'_t(x),
\end{equation}
where $\Gamma_5$ is given (for $N=6$) by
\begin{equation}
 \left( \Gamma_5 \right)_{st} = 
 \left( \begin{array}{cccccc} -\gamma_5 & 0 & 0 & 0 & 0 & 0 \\
                              0 & -\gamma_5 & 0 & 0 & 0 & 0 \\
                              0 & 0 & -\gamma_5 & 0 & 0 & 0 \\
                              0 & 0 & 0 & \gamma_5 & 0 & 0 \\
                              0 & 0 & 0 & 0 & \gamma_5 & 0 \\
                              0 & 0 & 0 & 0 & 0 & \gamma_5 
         \end{array} 
 \right)  \qquad (N=6).
\end{equation}
In particular, the light fermion transforms as 
Eq.~(\ref{eq:chiral-transformation-of-q}).
\[
\delta q(x) = \gamma_5 q(x), \quad
\delta \bar q(x) = \bar q(x) \gamma_5 .
\]
With this definition of the chiral transformation, 
the chiral symmetry is broken only by 
the diagonal $\frac{N}{2}$-th element of the hermitian
mass matrix Eq.~(\ref{eq:hermitian-mass-matrix}), 
i.e. the (diagonal) mass term of the $\frac{N}{2}$-th flavor:
\begin{equation}
\label{eq:chiral-property-domain-wall-D}
\left\{ \Gamma_5 D_{\rm DW}^\prime + D_{\rm DW}^\prime \Gamma_5 \right\}_{st}
=\frac{2}{a_5} \, 
 \gamma_5 \delta_{s \frac{N}{2}}\delta_{t \frac{N}{2}} .
\end{equation}

\subsection{Pauli-Villars field and truncated overlap Dirac operator}
In order to take the limit of infinite flavors 
and to relate the domain-wall fermion to the Dirac fermion described 
by the overlap Dirac operator, 
it is suitable to introduce, as a Pauli-Villars field, 
the $N$-flavor Wilson-Dirac bosons with the flavor-mixing 
mass matrix which is anti-periodic in the flavor 
space \cite{vranas-pauli-villars}:
\begin{eqnarray}
\label{eq:hermitian-mass-matrix-antiperiodic}
M^{\rm PV}_{st}&=&
\frac{1}{a_5}
\left( \begin{array}{cccccc}
           0 & 0 & 0 & 0 & -1& B\\
           0 & 0 & 0 & -1& B & 0 \\
           0 & 0 & -1& B & 0 & 0 \\
           0 &-1 & B & 0 & 0 & 0 \\
           -1& B & 0 & 0 & 0 & 0 \\ 
           B & 0 & 0 & 0 & 0 & 1
\end{array} \right) . \nonumber\\
\end{eqnarray}
Then the total action of the domain-wall fermion with the subtraction
of the Pauli-Villars field is given by
\begin{eqnarray}
\bar S_{\rm DW}
&=&\sum_{s,t=1}^N a^4 \sum_x
\bar \psi_{s}^\prime(x) 
\left\{ 
 \gamma_\mu \frac{1}{2}\left(\nabla_\mu+\nabla_\mu^\ast\right) \delta_{st} 
+ M_{st}^H \right\} \psi_{t}^\prime(x) 
\nonumber\\
&+&
\sum_{s,t=1}^N a^4 \sum_x
\bar \phi_{s}^\prime(x) 
\left\{ 
 \gamma_\mu \frac{1}{2}\left(\nabla_\mu+\nabla_\mu^\ast\right) \delta_{st} 
+ M^{\rm PV}_{st} 
\right\} \phi_{t}^\prime(x).  \nonumber\\
\end{eqnarray}
As shown by Neuberger in \cite{truncated-overlap} 
that the partition function of this total system at a finite 
flavor $N$ can be written by a single determinant of 
the truncated overlap Dirac operator:
\begin{equation}
\label{eq:subtracted-partition-function}
\bar Z_{\rm DW} = Z_{\rm DW} Z_{\rm PV} 
= \det \, a D_N .
\end{equation}
The truncated overlap Dirac operator is defined by 
Eq.~(\ref{eq:truncated-overlap-dirac-operator}) through 
the transfer matrix, which is given explicitly 
in the chiral basis \footnote{
The gamma matrices in the chiral basis are chosen as follows 
in our convention, 
\begin{equation}
  \gamma_\mu=\left( \begin{array}{cc} 0 & \sigma_\mu \\
                                      \sigma_\mu^\ast & 0 
                    \end{array} \right) , 
\quad 
  \gamma_5=\left( \begin{array}{cc} 1 & 0 \\
                                    0 & -1
                    \end{array} \right) ,  \quad
\sigma_\mu = \left( 1, \sigma_1, \sigma_2, \sigma_3 \right) .
\end{equation}
}
as 
\begin{equation}
T \equiv e^{- a_5 \bar H} = \left(
\begin{array}{cc} \frac{1}{B} & \frac{1}{B} C \\
                 -C^\dagger \frac{1}{B} 
& B + C^\dagger \frac{1}{B} C 
\end{array} \right),
\end{equation}
where
\begin{equation}
\label{eq:operator-C}
  C = a_5 \, \sigma_\mu \, 
\frac{1}{2}\left(\nabla_\mu+\nabla_\mu^\ast\right) 
\end{equation}
and $B$ is given by Eq.~(\ref{eq:operator-B}).
Thus the domain-wall fermion with the subtraction
of the Pauli-Villars field is equivalent to the Dirac fermion 
described by the truncated overlap Dirac operator:
\begin{equation}
\label{eq:truncated-overlap-Dirac-fermion}
S_N 
= a^4 \sum_x \bar \psi(x) \, D_N \, \psi(x) .
\end{equation}

The overlap Dirac operator, which is obtained from $D_N$ in the 
limit of the infinite flavors (but with a finite $a_5$),
is denoted as $\widetilde D \equiv \lim_{N\rightarrow \infty} D_N$. 
This reduces to $D$ of 
Eq.~(\ref{eq:truncated-overlap-dirac-operator}) in the limit that
$a_5$ vanishes.

\subsection{Effective action of the light fermion of domain-wall fermion}
\label{sec:effective-action-of-q-calculation}

Now we evaluate the effective action of the light fermion 
field by integrating out $N-1$ heavy flavors of the domain-wall fermion:
\begin{equation}
  S_N^{\rm eff} = a^4 \sum_x \bar q(x) \, D_N^{\rm eff} \, q(x). 
\end{equation}
As mentioned in the introduction,  this can be achieved 
by calculating the propagator of $q(x)$, because
it should be given by the inverse of the effective Dirac operator 
for the field variables $q(x)$ and $\bar q(x)$. 
In order to obtain the propagator of these fields, we introduce
the sources for them,
\begin{equation}
a^4 \sum_x \left\{  \bar J(x) q(x) +   \bar q(x) J(x) \right\} .
\end{equation}

We first describe the case with four flavors 
($N=4$) for simplicity and then generalize the result to any flavors. 
In the chiral basis of the gamma matrices,
the original action 
of the domain-wall fermion Eq.~(\ref{eq:action-domain-wall-fermion})
can be written in the matrix form as
\begin{equation}
  S_{\rm DW} = a^4 \sum_x \bar \Psi(x) D_{\rm DW} \Psi(x), \qquad (N=4)
\end{equation}
where
\begin{eqnarray}
\bar \Psi(x)&=& 
\left(\begin{array}{cccccccc} 
\bar \psi_{1L} & \bar \psi_{1R} &
\bar \psi_{2L} & \bar \psi_{2R} &
\bar \psi_{3L} & \bar \psi_{3R} &
\bar \psi_{4L} & \bar \psi_{4R} 
\end{array}\right),  \\
\nonumber\\
D_{\rm DW}&=& 
\frac{1}{a_5}
\left( \begin{array}{cccccccccc} 
B & C & 0 & 0 & 0 & 0 & 0 & 0\\
-C^\dagger & B & 0 & -1 & 0 & 0 & 0 & 0\\
-1 & 0 & B & C & 0 & 0 & 0 & 0\\
0 & 0 & -C^\dagger & B & 0 & -1 & 0 & 0\\
0 & 0 & -1 & 0 & B & C & 0 & 0\\
0 & 0 & 0 & 0 & -C^\dagger & B & 0 & -1\\
0& 0 & 0 & 0 & -1 & 0 & B & C \\
0& 0 & 0 & 0 & 0 & 0 & -C^\dagger & B \\
\end{array} \right) , \\ 
\nonumber\\
 \Psi(x) &=& \left(\begin{array}{c} 
\psi_{1R} \\ \psi_{1L} \\
\psi_{2R} \\ \psi_{2L} \\
\psi_{3R} \\ \psi_{3L} \\
\psi_{4R} \\ \psi_{4L} 
\end{array}\right) .
\end{eqnarray}
Following Neuberger \cite{truncated-overlap}, 
we then make the Dirac operator almost upper 
triangle. We first exchange the right-handed component 
and the left-handed component of each flavor in $\Psi(x)$:
\begin{equation}
\left(  \begin{array}{c} \psi_{t R} \\
                         \psi_{t L} \end{array} \right)
\Longrightarrow 
\left(  \begin{array}{c} \psi_{t L} \\
                         \psi_{t R} \end{array} \right)
= 
\left(\begin{array}{cc} 0 & 1 \\ 1 & 0 \end{array}\right) \, 
\left(  \begin{array}{c} \psi_{t R} \\
                         \psi_{t L} \end{array} \right),
\quad ( t = 1,2,3,4).
\end{equation}
Then the Dirac operator of the domain-wall fermion becomes
\begin{eqnarray}
D_{\rm DW} &\Longrightarrow& 
\frac{1}{a_5}
\left( \begin{array}{cccccccccc} 
C & B & 0 & 0 & 0 & 0 & 0 & 0\\
B & -C^\dagger  & -1 & 0 & 0 & 0 & 0 & 0\\
0 & -1 & C & B & 0 & 0 & 0 & 0\\
0 & 0 & B& -C^\dagger  & -1 & 0 & 0 & 0\\
0 & 0 & 0 & -1 & C & B & 0 & 0\\
0 & 0 & 0 & 0 & B & -C^\dagger & -1 & 0 \\
0& 0 & 0 & 0 & 0 & -1 & C & B \\
0& 0 & 0 & 0 & 0 & 0 & B & -C^\dagger \\
\end{array} \right) . \nonumber\\
\end{eqnarray}
We further move the first row down to the last row:
\begin{eqnarray}
&\Longrightarrow& D_{\rm DW}'' \equiv
\frac{1}{a_5}
\left( \begin{array}{cccccccccc} 
B & -C^\dagger  & -1 & 0 & 0 & 0 & 0 & 0\\
0 & -1 & C & B & 0 & 0 & 0 & 0\\
0 & 0 & B& -C^\dagger  & -1 & 0 & 0 & 0\\
0 & 0 & 0 & -1 & C & B & 0 & 0\\
0 & 0 & 0 & 0 & B & -C^\dagger & -1 & 0 \\
0& 0 & 0 & 0 & 0 & -1 & C & B \\
0& 0 & 0 & 0 & 0 & 0 & B & -C^\dagger \\
C & B & 0 & 0 & 0 & 0 & 0 & 0\\
\end{array} \right) . \nonumber\\
\end{eqnarray}
Accordingly, the components of the domain-wall fermion fields reads
\begin{eqnarray}
\bar \Psi(x)&\Longrightarrow& 
\bar \Psi'' (x)=
\left(\begin{array}{cccccccc} 
\bar \psi''_{1R} & \bar \psi''_{1L} 
& \bar \psi''_{2R} & \bar \psi''_{2L} 
& \bar \psi''_{3R} & \bar \psi''_{3L} 
& \bar \psi''_{4R} & \bar \psi''_{4L} 
\end{array}\right)  \nonumber\\
&& \phantom{\bar \Psi'' (x)} \equiv
\left(\begin{array}{cccccccc} 
\bar \psi_{1R} &
\bar \psi_{2L} & \bar \psi_{2R} &
\bar \psi_{3L} & \bar \psi_{3R} &
\bar \psi_{4L} & \bar \psi_{4R} & \bar \psi_{1L} 
\end{array}\right),  \nonumber\\
\\
\Psi(x) &\Longrightarrow& 
\Psi''(x) =
\left(\begin{array}{c} 
\psi''_{1L} \\ \psi''_{1R} \\ 
\psi''_{2L} \\ \psi''_{2R} \\ 
\psi''_{3L} \\ \psi''_{3R} \\ 
\psi''_{4L} \\ \psi''_{4R} 
\end{array}\right)  
\equiv
\left(\begin{array}{c} 
\psi_{1L} \\ \psi_{1R} \\ 
\psi_{2L} \\ \psi_{2R} \\ 
\psi_{3L} \\ \psi_{3R} \\ 
 \psi_{4L} \\ \psi_{4R} 
\end{array}\right) .
\end{eqnarray}
The sources for $q(x)$ and $\bar q(x)$ may be expressed in this 
upper-triangle basis as follows:
\begin{eqnarray}
&&  a^4 \sum_x \left\{ \bar J(x) 
\left[ 
 P_L \left(\begin{array}{cc} 0 & 1 \\ 1 & 0 \end{array}\right) 
 \psi''_{1}(x)
+P_R \left(\begin{array}{cc} 0 & 1 \\ 1 & 0 \end{array}\right) 
 \psi''_{N}(x) \right] 
\right.  \nonumber\\
&& \qquad\qquad \qquad\qquad \qquad\qquad \qquad\quad
\left.
+ \bar \psi''_{N} (x) 
\left(\begin{array}{cc} 0 & 1 \\ 1 & 0 \end{array}\right)
J(x) \right\} . \nonumber\\
\end{eqnarray}
In order to express the two by two blocks of $D''_{\rm DW}$, 
we introduce the following abbreviations 
\begin{eqnarray}
&&\alpha
= \frac{1}{a_5}
\left( \begin{array}{cc} B & -C^\dagger  \\
                                   0 & -1 \end{array} \right),
\quad
  \beta
= \frac{1}{a_5}
\left( \begin{array}{cc} 
-1 & 0 \\
C & B 
\end{array} \right),  \\
&& 
  \alpha_0
= \frac{1}{a_5}
\left( \begin{array}{cc} B & -C^\dagger  \\
                                   0 & 0 \end{array} \right), 
\quad
   \beta_0
= \frac{1}{a_5}
\left( \begin{array}{cc} 
0 & 0 \\
C & B 
\end{array} \right).
\end{eqnarray}

In this basis it is now easy to integrate fields from the first flavor down
to the last flavor.\footnote{ Note that $q_L(x)$  is 
still in the first component of $\Psi''(x)$, although 
$\bar q_L(x)$ is in the $N$-th component of $\bar \Psi''(x)$. 
This is why we could not evaluate the effective action directly in 
this basis which is convenient for the integration.}
The terms which include the field variables 
of the first flavor are following 
(The summation over the lattice indices $x$ with the measure factor $a^4$
is understood in the following equations.):
\begin{eqnarray}
&&
\left[  
  \bar J(x)  P_L \left(\begin{array}{cc} 0 & 1 \\ 1 & 0 \end{array}\right) 
+ \bar \psi''_N(x) \, \beta_0  
\right] \, \psi''_{1}(x)
\nonumber\\
&& \qquad\qquad\quad
+ \bar \psi''_1(x) \, \alpha \, \psi''_1(x)
\nonumber\\
&& \qquad\qquad\qquad\qquad\qquad
+ \bar \psi''_1(x) \, \beta \, \psi''_2(x) .
\end{eqnarray}
After integrating the first flavor, 
the terms which include the second flavor are found as follows:
\begin{eqnarray}
&&
\left[ 
  \bar J(x)  P_L \left(\begin{array}{cc} 0 & 1 \\ 1 & 0 \end{array}\right) 
+ \bar \psi''_N(x) \, \beta_0 
\right] 
\, \left( - \alpha^{-1} \beta \right) \psi''_2(x)
\nonumber\\
&&  \qquad\qquad\qquad
+ \bar \psi''_2(x) \, \alpha \, \psi''_2(x)
\nonumber\\
&& \qquad\qquad\qquad\qquad\qquad\quad
+ \bar \psi''_2(x) \, \beta \, \psi''_3(x) .
\end{eqnarray}
The integration of the second flavor leaves 
the terms which include the third flavor as follows:
\begin{eqnarray}
&&
\left[ 
  \bar J(x)  P_L \left(\begin{array}{cc} 0 & 1 \\ 1 & 0 \end{array}\right) 
+ \bar \psi''_N(x) \, \beta_0 
\right]
\, \left( - \alpha^{-1} \beta \right)^2 \psi''_3(x)
\nonumber\\
&&   \qquad\qquad\qquad
+ \bar \psi''_3(x) \, \alpha \, \psi''_3(x)
\nonumber\\
&& \qquad\qquad\qquad\qquad\qquad\quad
+ \bar \psi''_3(x) \, \beta \, \psi''_4(x) .
\end{eqnarray}
After the integration of the third flavor, only the forth flavor 
remains:
\begin{eqnarray}
\label{eq:last-flavor-terms}
&&  
\bar J(x) \left[
 P_L \left(\begin{array}{cc} 0 & 1 \\ 1 & 0 \end{array}\right) 
\left( - \alpha^{-1} \beta \right)^{N-1}
+P_R \left(\begin{array}{cc} 0 & 1 \\ 1 & 0 \end{array}\right) 
 \right] \psi''_N(x)
\nonumber\\
&& \qquad\qquad\qquad
+ 
\bar \psi''_N(x) \, 
\left[ \alpha_0
      +\beta_0 \, 
\left( - \alpha^{-1} \beta \right)^{N-1} \right] \psi''_N(x) 
\nonumber\\
&& \qquad\qquad\qquad\qquad\qquad\quad
+ \bar \psi''_N (x) 
\left(\begin{array}{cc} 0 & 1 \\ 1 & 0 \end{array}\right)
J(x), \qquad (N=4). \nonumber\\
\end{eqnarray}
As easily seen, this result holds true for any flavors $N$.

Now noting 
\begin{eqnarray}
\label{eq:alpha-beta-T}
\alpha_0 \alpha^{-1} &=& 
\left(\begin{array}{cc} 0 & 1 \\ 1 & 0 \end{array}\right) 
P_L 
\left(\begin{array}{cc} 0 & 1 \\ 1 & 0 \end{array}\right),  \nonumber\\
\beta_0 \beta^{-1} &=& 
\left(\begin{array}{cc} 0 & 1 \\ 1 & 0 \end{array}\right) 
P_R 
\left(\begin{array}{cc} 0 & 1 \\ 1 & 0 \end{array}\right),  \nonumber\\
\left(-\beta \, \alpha^{-1}\right) 
&=&
\left(\begin{array}{cc} 0 & 1 \\ 1 & 0 \end{array}\right) 
\gamma_5 \, e^{a_5 \widetilde H} \, \gamma_5 
\left(\begin{array}{cc} 0 & 1 \\ 1 & 0 \end{array}\right) ,
\end{eqnarray}
we can evaluate the factors of the first and second terms 
of Eq.~(\ref{eq:last-flavor-terms}) as follows:
\begin{eqnarray}
\label{eq:factor-1-2}
&&
\left[
 P_L \left(\begin{array}{cc} 0 & 1 \\ 1 & 0 \end{array}\right) 
\left( - \alpha^{-1} \beta \right)^{N-1}
+P_R \left(\begin{array}{cc} 0 & 1 \\ 1 & 0 \end{array}\right) 
 \right] \alpha^{-1}
\nonumber\\
&& \qquad \qquad = - a_5
\left[ P_R + P_L \, e^{N a_5 \widetilde H } \right] (-\gamma_5)
\left(\begin{array}{cc} 0 & 1 \\ 1 & 0 \end{array}\right) ,  \\
&& \left[ \alpha_0
      +\beta_0 \, 
\left( - \alpha^{-1} \beta \right)^{N-1} \right] \alpha^{-1}
\nonumber\\
&& \qquad \qquad 
= 
\left(\begin{array}{cc} 0 & 1 \\ 1 & 0 \end{array}\right) 
\left[ P_L + P_R \, e^{N a_5 \widetilde H } \right] (-\gamma_5)
\left(\begin{array}{cc} 0 & 1 \\ 1 & 0 \end{array}\right) .
\end{eqnarray}
From these results, we obtain
\begin{eqnarray}
  \left\langle q(x) \bar q(y) \right\rangle 
&=&  \frac{a_5}{a^4}
\left(P_R + P_L \, e^{N a_5 \widetilde H } \right)
\frac{1}{P_L + P_R \, e^{N a_5 \widetilde H}}
\nonumber\\
&=& \frac{a_5}{a^4}
\frac{1-\gamma_5 \tanh a_5 \frac{N}{2} \widetilde H }
     {1+\gamma_5 \tanh a_5 \frac{N}{2} \widetilde H } .
\end{eqnarray}
Then, the effective action of the light fermion turns out to be given
by the following effective Dirac operator:
\begin{eqnarray}
D_N^{\rm eff}(x,y) 
&=& 
\frac{1}{a^4}\left\langle q(x) \, \bar q(y) \right\rangle ^{-1} \nonumber\\
&=& \frac{1}{a_5} \, 
\frac{1+\gamma_5 \tanh  \frac{N}{2} a_5 \widetilde H}
     {1-\gamma_5 \tanh \frac{N}{2} a_5  \widetilde H} .
\end{eqnarray}

\subsection{The effective action of the light fermion and 
the Ginsparg-Wilson relation}

We may also consider the effective action of the $N$-th flavor of the 
bosonic Pauli-Villars field. As we see from 
Eq.~(\ref{eq:hermitian-mass-matrix-antiperiodic}), 
the difference between the mass matrices of the domain-wall fermion
and the Pauli-Villars field is only in the diagonal $N$-th 
element of the latter, which takes account of the anti-periodicity 
in the flavor space. Then the integration of the first $N-1$
components can be achieved just in the same way as the case of 
the domain-wall fermion. The effective action of the $N$-th flavor 
of the Pauli-Villars field turns out to be same as that of the 
light-fermion, but in this case, with the additional mass term. 
Thus the total effective action of the domain-wall fermion with the
subtraction can be written in the following form:
\begin{equation}
\label{eq:effective-action-subtracted}
  \bar S_N^{\rm eff}
= a^4 \sum_x \bar q(x) \, D_N^{\rm eff} \, q(x) 
+ a^4 \sum_x \bar Q(x) \left\{ D_N^{\rm eff}+\frac{1}{a_5} \right\} Q(x) ,
\end{equation}
where we have denoted the $N$-th flavor of the Pauli-Villars 
field by 
\begin{equation}
\label{eq:N-th-PV-field}
  Q(x) = \phi_N^\prime(x), \qquad \bar Q(x) = \bar \phi_N^\prime(x) .
\end{equation}
This result is consistent with 
Eq.~(\ref{eq:subtracted-partition-function}),
because the relation holds true
\begin{equation}
\label{eq:truncated-overlap-vs-effective-action-q}
\frac{ a_5 D_N^{\rm eff} }{1+ a_5 D_N^{\rm eff}}
= a D_N , 
\end{equation}
and the partition function calculated from the effective action
$\bar S_N^{\rm eff}$ ( Eq.~(\ref{eq:effective-action-subtracted})) 
is identical to that from the action with
the truncated overlap Dirac operator
$S_N$ ( Eq.~(\ref{eq:truncated-overlap-Dirac-fermion})):
\begin{eqnarray}
\bar Z_{\rm DW}
&=& \int [d q d \bar q] [d Q d \bar Q] 
             e^{- a^4 \sum_x \bar q(x) D_N^{\rm eff} q(x)
                - a^4 \sum_x \bar Q(x) \left\{ D_N^{\rm eff}
             +\frac{1}{a_5} \right\} Q(x) } \nonumber\\
&=& \int [d \psi d \bar \psi] 
             e^{- a^4 \sum_x \bar \psi(x) D_N \psi(x) } .
\end{eqnarray}

Eq.~(\ref{eq:truncated-overlap-vs-effective-action-q}) may be written
also in the following form:
\begin{eqnarray}
\label{eq:inverse-truncated-overlap-vs-propagator-q}
\frac{1}{a_5} \, { D_N^{\rm eff} }^{-1}
 + \delta(x,y)
&=& 
\frac{1}{a} \, { D_N^{\phantom{f}} }^{-1} .
\end{eqnarray}
Namely, the inverse of the effective Dirac operator 
(the propagator of the light fermion)
gives the inverse of the truncated overlap Dirac operator up to 
a local contact term.
This contact term just takes account of the 
chiral symmetry breaking in the Ginsparg-Wilson relation, 
which holds true for the overlap Dirac operator 
in the limit of the infinite flavors.
This result implies that the propagator of the light fermion 
reduces to 
{\it the chirally symmetric part of the inverse of the 
overlap Dirac operator} in the limit of the infinite flavors,
assuming that $\widetilde D$ does not have any zero mode:
\begin{equation}
\gamma_5 \, { D^{\rm eff} }^{-1}
+{ D^{\rm eff} }^{-1} \gamma_5 = 0  
\end{equation}
and
\begin{equation}
\label{eq:inverse-overlap-vs-propagator-q}
{\widetilde D}^{-1}
= 
\frac{a}{a_5} \, { D^{\rm eff} }^{-1}
 + a \delta(x,y) \qquad (N=\infty).
\end{equation}

It may be interesting to observe that 
the infinitely many flavors of the domain-wall fermion 
serve to prepare the chirally symmetric part of the inverse 
of the overlap Dirac operator and that 
it is the Pauli-Villars field ($N$-th flavor) that add it 
the contact term of the chiral symmetry breaking 
in the Ginsparg-Wilson relation.
This contrasts with the situation of the original derivation
of the Ginsparg-Wilson relation in \cite{ginsparg-wilson-rel}, 
where the chiral symmetry breaking is introduced 
by the kernel of the block-spin transformation. 

The properties of the chirally symmetric part of the inverse 
of the Dirac operator which satisfies the Ginsparg-Wilson relation
have been discussed extensively 
by Chiu et al. \cite{chiu-etal}. 
Eq.~(\ref{eq:truncated-overlap-vs-effective-action-q}) 
reduces to the expression of the Dirac operator in terms 
of the chirally symmetric part discussed there. 

As is also emphasized by these authors, 
the effective action of the light fermion, in the limit of infinite 
flavors, must necessarily be nonlocal. 
This can be easily checked in the free theory.
Then, the effective action of the light fermion itself
is not defined well and useful in this limit.
In fact, a subtlety appears when one attempts to calculate the axial
anomaly from the effective action, as will be discussed in the last
section. 

However, as we will discuss below, the light fermion field variables
$q(x)$ and $\bar q(x)$ still remain to be a good probe 
for the Dirac fermion described by the overlap Dirac operator.
They can be expressed by the Dirac fermion 
field $\psi(x)$ and $\bar \psi(x)$ through a certain local
expression including the overlap Dirac operator $\widetilde D$.
In this expression, we do not encounter any singularity associated 
with the nonlocal behavior of $D^{\rm eff}$. 
Moreover, the low energy observables of the domain-wall fermion
which are written in terms of $q(x)$ and $\bar q(x)$
can be expressed by $\psi(x)$ and $\bar \psi(x)$.
And these observables, in fact, turn out to have
good chiral properties from the point of view of the exact chiral 
symmetry based on the Ginsparg-Wilson relation.

\section{Chiral property of the light fermion}
\label{sec:chiral-property-q-and-psi}
\reseteqnum

With the result obtained in the previous section, 
we consider the relation between the light fermion field 
of the domain-wall fermion 
and the Dirac fermion field 
described by the truncated overlap Dirac operator.
Then we discuss the chiral property of the light fermion field
and various observables written by it
in view of the exact chiral symmetry based on the Ginsparg-
Wilson relation.

\subsection{Light fermion field $q(x)$ and $\bar q(x)$}

The correspondence between the light fermion field $q(x)$ and $\bar q(x)$
and the Dirac fermion field $\psi(x)$ and $\bar \psi(x)$ 
which is described by the truncated overlap Dirac operator 
is now easily understood. We may relate them by 
\begin{eqnarray}
\label{eq:relation-q-psi}
q(x)&=& Z \, \frac{1}{1+a_5 D_N^{\rm eff}} \, \psi(x) 
=Z \, \left( 1-\frac{a}{2} R D_N \right) \psi(x) , \nonumber\\
\bar q(x)&=& \bar \psi(x),
\end{eqnarray}
where 
$Z=\frac{a_5}{a}$ and $R=2$.
In the functional integral of the partition function,
the Jacobian of the change of the field variable from $q(x)$ 
to $\psi(x)$  along this relation,  
just compensates the determinant resulting from the integration 
of the last flavor of the bosonic Pauli-Villars field $Q(x)$.

A few comments are in order. 
As long as the equivalence of the partition functions
is concerned, there is an ambiguity 
in the correspondence Eq.~(\ref{eq:relation-q-psi}) 
which is related to the scale transformation:
\begin{equation}
  \psi(x) \longrightarrow z \, \psi(x), \qquad 
  \bar \psi(x) \longrightarrow z^{-1} \bar \psi(x).
\end{equation}
This scale factor may even depend on $D_N^{\rm eff}$.
We have fixed this freedom so that 
the chiral transformation of $\bar q(x)$ matches with that 
of $\bar \psi(x)$. As to the choice of the constant scale factor, 
it is our convention for simplicity.
The correspondence Eq.~(\ref{eq:relation-q-psi}) 
would be different if we adopt a different transformation 
of the exact chiral symmetry of L\"uscher from 
Eq.~(\ref{eq:chiral-transformation-of-psi}).

\subsection{Chiral transformation of the light fermion}
The explicit relation between $q(x)$ and $\psi(x)$ helps us 
to understand the chiral properties of the domain-wall
fermion in terms of the (would-be) exact chiral symmetry 
of the Ginsparg-Wilson fermion described by the (truncated) 
overlap Dirac operator.
First of all, 
Eq.~(\ref{eq:inverse-truncated-overlap-vs-propagator-q}) 
relates quantitatively the chiral symmetry breaking in the domain-wall
fermion to the breaking of the Ginsparg-Wilson relation 
in the truncated overlap Dirac fermion:
these breakings can be characterized by a single quantity
\begin{eqnarray}
\label{eq:breaking-delta}
\delta_N &\equiv& 
Z^{-1}\left(
\gamma_5 {D_N^{\rm eff} }^{-1} 
+ { D_N^{\rm eff} }^{-1}  \gamma_5   \right) \\
&\equiv& 
\gamma_5 { D_N }^{-1}  +{ D_N }^{-1}  \gamma_5 - a R \gamma_5 . 
\end{eqnarray}

We may consider the chiral transformation
for the Dirac fermion described by the truncated overlap
Dirac operator a l\'a L\"uscher 
\begin{equation}
\label{eq:chiral-transformation-of-psi-truncated}
  \delta \psi(x) = \gamma_5 \left( 1-aRD_N \right) \psi(x), \quad
  \delta \bar \psi(x) = \bar \psi(x) \gamma_5 .
\end{equation}
The chiral symmetry of the action is broken by the amount
\begin{equation}
  \delta S_N = a^4 \sum_x \bar \psi(x) \Delta_N \psi(x),
\end{equation}
\begin{equation}
\label{eq:breaking-Delta}
\Delta_N \equiv 
\gamma_5 \left\{ 1- \left(\tanh \frac{N}{2} a_5 \widetilde H \right)^2
\right\}
= D_N \delta_N D_N .
\end{equation}
This breaking vanishes in the limit of infinite flavors $N=\infty$, 
as long as the eigenvalues of $\widetilde H$ 
are bounded from zero uniformly with respect to 
the gauge fields in consideration. 
We assume this in the following discussions.
We also assume that $\widetilde D$, 
the overlap Dirac operator in the limit of the infinite flavors, 
should not have any zero mode and that its inverse should exist,
in order to assure that $\delta_N$ should vanish 
in this limit, too.
When we discuss the anomaly of the domain-wall fermion
in relation to the index of $\widetilde D$, we introduce 
the mass term of the light fermion to assure the limit.

The chiral transformation of 
Eq.~(\ref{eq:chiral-transformation-of-psi-truncated}) 
induces the transformation of the light fermion as follows:
\begin{eqnarray}
\delta q(x) 
&=& Z \, \left( 1-\frac{a}{2} R D_N \right) \delta \psi(x) 
\nonumber\\
&=& Z \, \left( 1-\frac{a}{2} R D_N \right)\,  
\, \gamma_5 \left( 1-a R D_N \right)\,  \psi(x) 
\nonumber\\
&=& Z \, 
\left\{ 
\gamma_5 \left( 1-a R D_N \right) 
-\frac{a}{2} R D_N \, \gamma_5 \left( 1-a R D_N \right)
\right\} \,  \psi(x) 
\nonumber\\
&=& Z \, 
\left\{ 
\gamma_5 \left( 1-\frac{a}{2} R D_N \right) 
- \frac{a}{2} R \Delta_N
\right\} \,  \psi(x) 
\nonumber\\
&=& 
\gamma_5 \, q(x)- a_5 \Delta_N \,  \psi(x) .
\end{eqnarray}
And it reduces to the transformation of 
Eq.~(\ref{eq:chiral-transformation-of-q}) in the limit
of the infinite flavors.
Thus we can see how the chiral transformation 
of Eq.~(\ref{eq:chiral-transformation-of-q}), 
which is adopted by Furman and Shamir 
(See also Eq.~(\ref{eq:chiral-transformation-of-DW})), is related 
to that of L\"uscher Eq.~(\ref{eq:chiral-transformation-of-psi}) 
in the limit of the infinite flavors.

\subsection{Low energy observables in terms of the light fermion}

Next we consider the low energy observables of the domain-wall fermion, 
which are written in terms of $q(x)$ and $\bar q(x)$.
We discuss the relation to the observables in terms of the Dirac
fermion field which is described by the overlap Dirac operator and 
whose chiral property is governed by the Ginsparg-Wilson relation.

\subsubsection{Scalar and pseudo scalar bilinear operators}
First of all, from the explicit relation 
Eq.~(\ref{eq:relation-q-psi}) between the light 
fermion field and the Dirac field described by the truncated
overlap Dirac operator, 
we obtain the relation of the scalar and pseudo scalar bilinear operators:
\begin{eqnarray}
\bar q(x) q(x) 
&=& Z \, \bar \psi(x) \left(1- \frac{a}{2}RD_N \right)\psi(x) , \\
\bar q(x) \gamma_5 q(x) 
&=& Z \, \bar \psi(x) \gamma_5 \left(1- \frac{a}{2}RD_N \right)\psi(x) .
\end{eqnarray}
In this respect, it is interesting to note that the scalar and pseudo 
scalar bilinear operators in the r.h.s.
consist the exact chiral multiplet of the chiral transformation 
Eq.~(\ref{eq:chiral-transformation-of-psi}) 
in the limit of infinite flavors , 
as discussed by Niedermayer \cite{neidermayer-lat98}.
In fact, these operators can be written in the chiral components
defined by $\hat \gamma_5 = \gamma_5\left(1-a R \widetilde D\right)$ for
$\psi(x)$ and by $\gamma_5$ for $\bar \psi(x)$ as follows:
\begin{eqnarray}
\bar q(x) q(x) &\longrightarrow&
\bar \psi_L(x) \psi_R(x) + \bar \psi_R(x) \psi_L(x) , \\
\bar q(x) \gamma_5 q(x) &\longrightarrow&
\bar \psi_L(x) \psi_R(x) - \bar \psi_R(x) \psi_L(x) .
\end{eqnarray}

\subsubsection{Conserved vector currents}
The vector current of the domain-wall fermion is conserved
at a finite flavor $N$. It is expected to correspond to 
the conserved vector current of the truncated overlap Dirac fermion.
We will show that the vector current of the domain-wall fermion,
if it is probed by the light fermion field $q(x)$ and $\bar q(x)$ 
at low energy, just corresponds to 
the conserved vector current of the truncated overlap Dirac fermion.

In general, a vector current may be defined 
with the kernel 
\begin{equation}
  V^a_\mu(x) = \sum_{y,z} \bar \psi(x) K^a_\mu(x;y,z) \psi(x) .
\end{equation}
The kernel is obtained from the Dirac operator by
introducing the auxiliary vector field
\begin{equation}
U^B_\mu(x)= \exp \left( i B^a_\mu(x) T^a  \right), 
\end{equation}
where $T^a$ are assumed as the generators of 
the $U(N_F)$ real flavor group,
and by differentiating 
the Dirac operator with respect to the vector field: 
\begin{equation}
\delta D(y,z) = \sum_x B^a_\mu(x) \, K^a_\mu(x;y,z) 
+ {\cal O}\left(B^2\right).
\end{equation}
By this procedure, we can obtain 
the kernel for the domain-wall fermion
from $D_{\rm DW}^\prime
\equiv\gamma_\mu\frac{1}{2}\left(\nabla_\mu+\nabla_\mu^\ast\right)+
M^{\rm H} $, which we denote as $K^a_{\mu {\rm DW}}$. 
The kernel for the truncated overlap Dirac fermion
is obtained from $D_N$, which we denote as $K^a_{\mu N}$.
Then, the vector current of the domain-wall fermion with
the subtraction can be written as 
\begin{equation}
V^a_{\mu {\rm DW}}(x) 
= \sum_{st}^N \bar \psi^\prime_s 
  \left\{ K^a_{\mu {\rm DW}}(x) \right\}_{st} \psi^\prime_t 
.
\end{equation}
The vector current of the truncated overlap Dirac fermion can be
written as
\begin{equation}
  V^a_N(x) = \bar \psi \, K^a_{\mu N}(x) \, \psi.
\end{equation}

For these two vector currents, we can infer the following identity:
\begin{equation}
\label{relation-of-vector-currents-q}
  \left\langle q(y) \, V^a_{\mu {\rm DW}}(x) \, 
\bar q(z) \right\rangle_c
= Z \left\langle \psi(y) \, V^a_{\mu N}(x) \, \bar \psi(z) \right\rangle_c.
\end{equation}
This identity follows from the relation between $D_{\rm DW}^{-1}$ and 
$D_N^{-1}$ given by 
Eq.~(\ref{eq:inverse-truncated-overlap-vs-propagator-q}):
in fact, it follows that 
\begin{equation}
\label{eq:inverse-truncated-overlap-vs-propagator-q-variation}
\left\{ \delta  {D_{\rm DW}'}^{ -1} \right\}_{NN} 
= \delta { D_N^{\rm eff} }^{-1}
= Z \delta D_N^{-1} .
\end{equation}
Using the identity $ \delta D^{-1} = - D^{-1} \delta D D^{-1}$, 
it can be expressed as Eq.~(\ref{relation-of-vector-currents-q}).

\subsubsection{Almost conserved axial vector currents}

According to Shamir and Furman \cite{boundary-fermion-QCD}, 
the axial vector current of the domain-wall fermion
is conserved up to the corrections suppressed exponentially 
in the number of flavors. Then it is expected that 
this axial vector current is related in the limit of the infinite 
flavors to the conserved axial vector current of the overlap
Dirac fermion. 
We will show that the axial vector current of the domain-wall fermion,
if it is probed by the light fermion field $q(x)$ and $\bar q(x)$ 
at low energy, reduces to 
the conserved axial vector current of the overlap Dirac fermion.

The axial vector current of the domain-wall fermion
is defined with the kernel of the vector current as
\begin{equation}
A^a_{\mu {\rm DW}}(x) 
= 
\sum_{s,t}^N 
\bar \psi_s^\prime
\left\{ K^a_{\mu {\rm DW}} (x) \, \Gamma_5  \right\}_{st} \psi_t^\prime .
\end{equation}
On the other hand, the axial vector current of the 
truncated overlap Dirac fermion may be defined as 
\begin{equation}
 A^a_{\mu N}(x) = 
\bar \psi \, 
K^a_{\mu N} (x) \, \gamma_5\left(1-aRD_N\right) \, \psi ,
\end{equation}
which naturally follows from the chiral transformation 
Eq.~(\ref{eq:chiral-transformation-of-psi-truncated}). 
This current reduces to the Noether current associated 
with the exact chiral symmetry of L\"uscher \cite{axial-current} 
in the limit of the infinite flavors, which we denote as 
\begin{equation}
\widetilde A^a_{\mu}(x) = 
\bar \psi \, 
\widetilde K^a_{\mu } (x) \, 
\gamma_5\left(1-aR \widetilde D\right) \, \psi  .
\end{equation}

For these two axial vector currents, we can infer the following
relation:
\begin{equation}
\label{relation-of-axial-vector-currents-q}
\lim_{N\rightarrow \infty}
\left\langle q(y) \, A^a_{\mu {\rm DW}}(x) \, \bar q(z) 
\right\rangle_c
= 
Z \left\langle \psi(y) \, \widetilde A^a_{\mu }(x) \, 
\bar \psi(z) \right\rangle_c . 
\end{equation}
This relation can be shown in the following way.
The Dirac operator of the domain-wall fermion satisfies
\begin{equation}
\left\{ \Gamma_5 D_{\rm DW}^\prime + D_{\rm DW}^\prime \Gamma_5 \right\}_{st}
=2 \gamma_5 \delta_{s \frac{N}{2}}\delta_{t \frac{N}{2}} .
\end{equation}
On the other hand, the truncated overlap Dirac operator satisfies
\begin{equation}
\label{eq:chiral-property-truncated-overlap-D}
  \gamma_5 D_N + D_N \gamma_5 \left(1- a R D_N \right) = \Delta_N .
\end{equation}
Then it turns out that 
the vector and axial vector currents are related each other as follows:
\begin{eqnarray}
\label{relation-of-currents-domainwall}
&& 
\left\langle q(y) \, A^a_{\mu {\rm DW}}(x) \, \bar q(z) \right\rangle
+\left\langle q(y) \, V^a_{\mu {\rm DW}}(x) \, \bar q(z) \right\rangle
\gamma_5 
\nonumber\\
&&=
\sum_s^N \left\{ {D_{\rm DW}}^{-1} K^a_{\mu {\rm DW}}(x) \right\}_{N,s} \cdot 
\left\{ {D_{\rm DW}}^{-1} \right\}_{s,\frac{N}{2}} 2 \gamma_5 
\left\{ {D_{\rm DW}}^{-1} \right\}_{\frac{N}{2},N}(y,z) , \nonumber\\
\end{eqnarray}
and 
\begin{eqnarray}
\label{relation-of-currents-truncated-overlap}
&&
\left\langle \psi(y) \, A^a_{\mu N}(x) \, \bar \psi(z) \right\rangle 
+\left\langle \psi(y) \, V^a_{\mu N}(x) \, \bar \psi(z)\right\rangle
 \gamma_5
\nonumber\\
&& \qquad\qquad\qquad
=
\left\{ {D_N}^{-1} K_{\mu N}(x) \right\}
{D_N}^{-1} \Delta_N {D_N}^{-1}(y,z) .
\end{eqnarray}
Using Eq.~(\ref{relation-of-vector-currents-q}), we obtain
\begin{eqnarray}
\label{relation-of-axial-vector-currents-q-N}
&& \left\langle q(y) \, A^a_{\mu {\rm DW}}(x) \, \bar q(z) \right\rangle
\nonumber\\
&&\quad = 
Z \left\langle \psi(y) \, A^a_{\mu N}(x) \, \bar \psi(z) \right\rangle
-Z {D_N}^{-1} K_{\mu N}(x) {D_N}^{-1} \Delta_N {D_N}^{-1}(y,z) . 
\nonumber\\
&& \qquad
+
\left\{ {D_{\rm DW}}^{-1} K_{\mu {\rm DW}}(x) {D_{\rm DW}}^{-1} 
\right\}_{N,\frac{N}{2}} 2 \gamma_5 
\left\{ {D_{\rm DW}}^{-1} \right\}_{\frac{N}{2},N} (y,z) . 
\nonumber\\
\end{eqnarray}

Now, it is possible to evaluate 
the propagator between the $\frac{N}{2}$-th flavor and the $N$-th 
flavor of the domain-wall fermion, 
using the similar method to evaluate the propagator of the light
fermion in section~\ref{sec:effective-action-of-q}. 
The calculation is described in the 
appendix~\ref{sec:propagators-heavy-modes}.
The result can be expressed as
\begin{eqnarray}
\left\{ D_{\rm DW} \right\}^{-1}_{\frac{N}{2} N} (x,y) 
&\equiv& 
\left\langle \psi_{\frac{N}{2}}^\prime(x) \, \bar q(y) \right\rangle
\nonumber\\
&=& \frac{1}{2 \cosh \frac{N}{2} a_5 \widetilde H } \, D_N^{-1}(x,y) .
\end{eqnarray}
From this and Eq.~(\ref{eq:breaking-Delta}), 
we infer that the second and third terms of the right-hand sides of
Eq.~(\ref{relation-of-axial-vector-currents-q-N}) 
vanish in the limit of the infinite flavors 
and we finally obtain 
Eq.~(\ref{relation-of-axial-vector-currents-q}). 

Thus the vacuum expectation value of the axial vector current of
the domain-wall fermion with respect to the light fermion field
leads directly to that of the axial vector current associated with
the exact chiral symmetry based on the Ginsparg-Wilson relation.
Note that our results hold true with dynamical gauge fields 
as long as the eigenvalues of $\widetilde H$ 
are bounded from zero uniformly with respect to the gauge fields. 
The renormalization factor of the axial current 
of the domain-wall fermion thus reduces to unity in the limit 
of the infinite flavors. Our result is consistent with 
the perturbative result at one-loop 
obtained by Aoki and Taniguchi \cite{aoki-taniguchi}.

\subsubsection{Correspondence of various observables}

From the above results, we can obtain the 
correspondence of the various observables of the light fermion 
of the domain-wall fermion to the observables of 
the Dirac fermion which is described by the (truncated) overlap
Dirac operator. We summarize it here.
(Note again that in this paper, the bra-ket symbol $\langle \cdots 
\rangle$ denotes the Wick contraction of the fermion fields in it
by their propagators, not the fermionic vacuum expectation values
which must includes the weight of the fermion action.
The bra-ket symbol with the subscript $c$ denotes 
the connected contraction. )

\begin{itemize}

\item Correlation of the vector currents:
\begin{equation}
\left\langle V^a_{\mu {\rm DW}}(x) \,\, 
\bar q(y) \gamma_\nu T^b q(y) \right\rangle_c
= Z \left\langle V^a_{\mu N}(x) \,\, 
\bar \psi(y) \gamma_\nu T^b \psi(y) 
\right\rangle_c .
\end{equation}

\item Correlation of the pseudo scalar densities:
\begin{eqnarray}
&&\left\langle \bar q(x)\gamma_5 T^a q(x) \, \, 
               \bar q(y)\gamma_5 T^b q(y) 
  \right\rangle \nonumber\\
&& \quad = 
Z^2\left\langle 
\bar \psi(x)\gamma_5\left(1-\frac{a}{2}RD_N\right) T^a \psi(x) 
\, \, 
\bar \psi(y)\gamma_5\left(1-\frac{a}{2}RD_N\right) T^b \psi(y) 
 \right\rangle . \nonumber\\
\end{eqnarray}

\item Correlation of the axial vector current and pseudo scalar density:
\begin{equation}
\lim_{N\rightarrow \infty}
\left\langle A^a_{\mu {\rm DW}}(x) \,\, 
\bar q(y) \gamma_5 T^b q(y) \right\rangle_c
= Z \left\langle \widetilde A^a_{\mu }(x) \,\, 
\bar \psi(y) \gamma_5 T^b \psi(y) 
\right\rangle_c .
\end{equation}

\item Amplitude of $K^0$-$\bar K^0$ mixing:
\begin{eqnarray}
&& \lim_{N\rightarrow \infty}
\left\langle \bar d_N s_N(y) 
\left( \sum_{st}^N \, \bar s_s 
\left\{ K_{\mu {\rm DW}} (x) \left(1- \Gamma_5 \right) \right\}_{st} 
          d_t  \, \right)^2 
             \bar d_N s_N (z) \right\rangle
\nonumber\\
&& \qquad
= Z^2
\left\langle \bar d s(y) 
\left( \, \bar s \, 
\widetilde K_\mu(x) \left(1- \hat \gamma_5 \right) d  \, \right)^2 
             \bar d s(z) \right\rangle   ,
\end{eqnarray}
where $s_t(x)$ and $d_t(x)$ $(t=1,\cdots,N)$ are the 
domain-wall fermion fields for s-quark and d-quark, respectively.
$s(x)$ and $d(x)$ are the Dirac fermion fields 
which are described by the overlap Dirac operator $\widetilde D$
for s-quark and d-quark, respectively.

\end{itemize}

Thus we see that
the observables in the light fermion field variables 
$q(x)$ and $\bar q(x)$ of the domain-wall fermion 
(with the subtraction of the Pauli-Villars field) 
leads directly to the observables 
which have good chiral property with 
respect to the exact chiral symmetry based on 
the Ginsparg-Wilson relation.

\section{Axial anomaly of domain-wall fermion}
\label{sec:axial-anomaly-q-and-psi}
\reseteqnum

The explicit chiral symmetry breaking 
in the domain-wall fermion under 
the chiral transformation Eq.~(\ref{eq:chiral-transformation-of-DW})
is expected to reproduce
the axial anomaly \cite{axial-anomaly-in-domain-wall}.
Here we examine this chiral symmetry breaking term
in relation to the axial anomaly 
which is associated with the exact
chiral symmetry based on the Ginsparg-Wilson relation
\cite{index-theorem-at-finite-lattice,exact-chiral-symmetry},
\begin{equation}
\label{eq:anomaly-of-GW-fermion}
  - a {\rm tr} \gamma_5 R \widetilde D(x,x) .
\end{equation}

\subsection{Axial anomaly}

With the definition of
the chiral transformation Eq.~(\ref{eq:chiral-transformation-of-DW}),
the chiral symmetry breaking occurs at 
the diagonal 
$\frac{N}{2}$-th element of the hermitian mass matrix 
Eq.~(\ref{eq:hermitian-mass-matrix}), i.e. 
mass term of the $\frac{N}{2}$-th flavor.
In fact,
the flavor singlet axial vector current of the domain-wall fermion
satisfies the axial Ward-Takahashi identity:
\begin{equation}
\label{eq:axial-WT-identity-DW}
  \partial_\mu^\ast \left\langle A_{\mu {\rm DW}}(x) \right\rangle 
= 
\frac{2}{a_5}
\left\langle \bar \psi^\prime_{\frac{N}{2}}(x) \, \gamma_5 \, 
\psi^\prime_{\frac{N}{2}}(x) \right\rangle .
\end{equation}
It is expected that the explicit breaking in the r.h.s.
of the above identity should reproduce the axial anomaly.
It has been shown by a perturbative calculation that 
it is indeed the case \cite{axial-anomaly-in-domain-wall}.

It is actually possible to evaluate the breaking term
non-perturbatively.
Using the similar method to evaluate the propagator of the light
fermion in section~\ref{sec:effective-action-of-q}, the 
propagator of the $\frac{N}{2}$-th flavor fermion 
of the domain-wall fermion is evaluated explicitly as 
\begin{eqnarray}
\label{eq:chiral-symmetry-breaking-domain-wall-fermion}
  \left\langle \psi^\prime_{\frac{N}{2}}(x) \, 
\bar \psi^\prime_{\frac{N}{2}}(y) 
  \right\rangle 
&=& \frac{a_5}{a^4} \left(-1 + \gamma_5 \frac{1}{2} a R D_N \gamma_5 
\right. \nonumber\\
&& \qquad\qquad \left.
  +\frac{1}{2a} \frac{1}{\cosh \frac{N}{2} a_5 \widetilde H } 
   D_N^{-1} \gamma_5 \frac{1}{\cosh \frac{N}{2} a_5 \widetilde H }
  \gamma_5  \right) . \nonumber\\
\end{eqnarray}
The calculation is described in the 
appendix~\ref{sec:propagators-heavy-modes}.
Then it follows immediately that 
the chiral symmetry breaking term can be written as 
\begin{eqnarray}
\frac{2}{a_5} 
\left\langle \bar \psi^\prime_{\frac{N}{2}}(x) 
\gamma_5 \psi^\prime_{\frac{N}{2}}(x)
  \right\rangle
&=& \frac{1}{a^4} 
\left( - a {\rm tr} \gamma_5 R D_N (x,x)  
\phantom{
 D_N^{-1} \gamma_5 \frac{1}{\cosh^2 \frac{N}{2} a_5 \widetilde H}(x,x) 
}
\right.\nonumber\\
&& \qquad \qquad \left. 
- {\rm tr}
 D_N^{-1} \gamma_5 \frac{1}{a} 
                   \frac{1}{\cosh^2 \frac{N}{2} a_5 \widetilde H}(x,x) 
\right)
\nonumber\\
&=& \frac{1}{a^4} \left(
- a{\rm tr} \gamma_5 R D_N (x,x) - {\rm tr} D_N^{-1}  \Delta_N (x,x)
\right) .
\nonumber\\
\end{eqnarray}
We can see that it reduces to the anomaly 
Eq.~(\ref{eq:anomaly-of-GW-fermion}) in the limit of the infinite flavors.

\subsection{Relation to the index of overlap Dirac operator}

We next discuss the relation of the axial anomaly 
of the domain-wall fermion to 
the index of the overlap Dirac operator $\widetilde D$.
For this purpose, we consider the situation
so that $\widetilde D$ has zero modes. 
In order to make the limit of the infinite flavors well-defined, 
we introduce the mass term of the light fermion.
\begin{equation}
  m \bar q(x) q(x) .
\end{equation}

Then it follows from the axial Ward-Takahashi identity that
\begin{equation}
\label{eq:axial-WT-ideinty-DW-mqq}
a^4 \sum_x 2 \left\langle \bar \psi^\prime_{\frac{N}{2}}(x) 
\gamma_5 \psi^\prime_{\frac{N}{2}}(x)
\right\rangle
+ a^4 \sum_x 2 m  \left\langle \bar q (x) \gamma_5 q(x)  \right\rangle = 0 .
\end{equation}
The second term of the l.h.s. is evaluated as 
\begin{eqnarray}
\label{eq:anomalous-term-q}
- a^4 \sum_x 2 m  \left\langle \bar q (x) \gamma_5 q(x)  \right\rangle 
&=& 2 m {\rm Tr} \gamma_5 \frac{1}{ D_N^{\rm eff} + m }  \nonumber\\
&=& 2 m {\rm Tr} \gamma_5 \left(1-\frac{a}{2} R D_N \right) 
\frac{1}{ D_N + m \left(1-\frac{a}{2} R D_N \right) } \nonumber\\
&=& - a {\rm Tr} \gamma_5 R D_N \nonumber\\
&& - (1-a m) {\rm Tr} \Delta_N  
   \frac{1}{ D_N + m \left(1-\frac{a}{2} R D_N \right) } . \nonumber\\
\end{eqnarray}
Therefore we obtain in the limit of the infinite flavors 
\begin{equation}
\label{eq:anomalus-term-DW-and-index}
\lim_{N \rightarrow \infty}
a^4 \sum_x 2 \left\langle \bar \psi^\prime_{\frac{N}{2}}(x) 
\gamma_5 \psi^\prime_{\frac{N}{2}}(x)
\right\rangle
= -a {\rm Tr} \gamma_5 R \widetilde D  
= 2 \, {\rm Index} \left(\widetilde D \right).  
\end{equation}
Since all the dependence on $m$ comes with $\Delta_N$, 
this limit does not depend on the value of $m$.

This explicit chiral symmetry breaking term has been 
evaluated numerically 
by Argonne group \cite{lagae-sinclair}
as the probe of the topological charge.
We have seen how this quantity is related to 
the index of the overlap Dirac operator which 
satisfies the Ginsparg-Wilson relation.

\section{Contributions of the Pauli-Villars field}
\label{sec:Pauli-Villars-field}
\reseteqnum

Finally, we will discuss the contributions of the Pauli-Villars
field to vector and axial vector currents and axial anomaly. 
In the previous sections, we have discussed 
the currents of the domain-wall fermion 
probed by the light fermion field $q(x)$ and $\bar q(x)$.
However, if we consider the currents themselves and 
their correspondence to those of the overlap Dirac fermion, 
we need to include the contribution of the Pauli-Villars fields.

\subsection{Vector and axial vector currents}
The conserved vector current of the domain-wall fermion 
with the subtraction of the Pauli-Villars fields is defined by
\begin{equation}
\overline{V}^a_{\mu {\rm DW}}(x) 
= \sum_{st}^N \bar \psi^\prime_s 
  \left\{ K^a_{\mu {\rm DW}}(x) \right\}_{st} \psi^\prime_t 
+ \sum_{st}^N \bar \phi^\prime_s 
  \left\{ K^a_{\mu {\rm DW}}(x) \right\}_{st} \phi^\prime_t .
\end{equation}
Note that we obtain the same kernel for the Pauli-Villars field 
as the domain-wall fermion, 
from $D_{\rm PV}^\prime
\equiv\gamma_\mu\frac{1}{2}\left(\nabla_\mu+\nabla_\mu^\ast\right)+
M^{\rm PV}$.
The almost conserved axial vector current is defined similarly by
\begin{equation}
\overline{A}^a_{\mu {\rm DW}}(x) 
= 
\sum_{s,t}^N 
\bar \psi_s^\prime
\left\{ K^a_{\mu {\rm DW}} (x) \, \Gamma_5  \right\}_{st} \psi_t^\prime 
+\sum_{s,t}^N 
\bar \phi_s^\prime
\left\{ K^a_{\mu {\rm DW}} (x) \, \Gamma_5  \right\}_{st} \phi_t^\prime .
\end{equation}

For these vector and axial vector currents, we can infer the following 
identities:
\begin{eqnarray}
\label{relation-of-vector-currents}
\langle \overline{V}^a_{\mu {\rm DW}}(x) \rangle 
&=&  \langle V^a_{\mu N}(x) \rangle , \\
\label{relation-of-axial-vector-currents}
\lim_{N\rightarrow \infty}
\left\langle \overline{A}^a_{\mu {\rm DW}}(x) \right\rangle
&=& 
\left\langle \widetilde A^a_{\mu }(x) \right\rangle .
\end{eqnarray}
The identity of the vector currents follows immediately from 
Eq.~(\ref{eq:subtracted-partition-function}). But we will show it
here directly for later use.
The l.h.s. of Eq.~(\ref{relation-of-vector-currents}), if 
multiplied with the auxiliary vector field $B^a_\mu(x)$,  can be 
rewritten as follows:
\begin{eqnarray}
\label{eq:derivation-relation-of-vector-currents}
({\rm l.h.s.})&\simeq & a^4 \sum_x B^a_\mu(x)  
\langle \overline{V}^a_{\mu {\rm DW}}(x) \rangle 
\nonumber\\
&=& - \sum_t^N {\rm tr} 
\left\{ \delta D'_{\rm DW} 
\left( {D'_{\rm DW}}^{-1}- {D'_{\rm PV}}^{-1}\right) \right\}_{tt} 
\nonumber\\
&=& - \sum_t^N {\rm tr} 
\left\{ \delta D'_{\rm DW} 
{D'_{\rm DW}}^{-1} \left(M^{\rm PV}-M^{\rm H}\right){D'_{\rm PV}}^{-1} \right\}_{tt} 
\nonumber\\
&=& \sum_t^N {\rm tr} 
\left\{ \left[D'_{\rm PV} -\left(M^{\rm PV}-M^{\rm H}\right)\right]
\delta {D'_{\rm DW}}^{-1} \left(M^{\rm PV}-M^{\rm H}\right){D'_{\rm PV}}^{-1} \right\}_{tt} 
\nonumber\\
&=& {\rm tr} 
\left\{ \delta {D'_{\rm DW}}^{-1} \right\}_{NN}
\frac{1}{a_5} \left\{1 - \frac{1}{a_5}{D'_{\rm PV}}^{-1} \right\}_{NN} .
\end{eqnarray}
In the last equality, we have noted that 
\begin{equation}
\left\{M^{\rm PV}-M^{\rm H}\right\}_{st}= \frac{1}{a_5} \delta_{sN}\delta_{Nt}.
\end{equation}

The propagator of the $N$-th flavor of the Pauli-Villars field
has been evaluated in section \ref{sec:effective-action-of-q} as
\begin{equation}
\frac{1}{a_5} \left\{ {D'_{\rm PV}}^{-1} \right\}_{NN} 
= \frac{a^4}{a_5} \langle Q(x) \bar Q(y) \rangle 
= \frac{1}{a_5 D_N^{\rm eff}+ 1 }
= \left( 1- a D_N \right) .
\end{equation}
With this result and 
Eq.~(\ref{eq:inverse-truncated-overlap-vs-propagator-q-variation}), 
we obtain 
\begin{eqnarray}
({\rm l.h.s.})
&=& {\rm tr} \delta {D_N}^{-1} \, D_N  \nonumber\\
&=& a^4 \sum_x B^a_\mu(x)  \, \langle V^a_{\mu N}(x)  \rangle  .
\end{eqnarray}

As to the identity of the axial vector current, 
using
Eqs.~(\ref{eq:chiral-property-domain-wall-D}) and
(\ref{eq:chiral-property-truncated-overlap-D}) and 
through the similar calculation as in 
Eq.~(\ref{eq:derivation-relation-of-vector-currents}) ,
we obtain
\begin{eqnarray}
\label{relation-of-axial-vector-currents-N}
&& \langle \overline{A}^a_{\mu {\rm DW}}(x) \rangle    \nonumber\\
&& \quad
= 
\langle A^a_{\mu N}(x) \rangle   
+ {\rm tr} K^a_{\mu N}(x) D_N^{-1} \Delta_N D_N^{-1} 
\nonumber\\
&& \qquad
- {\rm tr}
\left\{ {D_{\rm DW}}^{-1} K^a_{\mu {\rm DW}}(x) {D_{\rm DW}}^{-1}
\right\}_{N,\frac{N}{2}}
 2 \gamma_5 
\left\{ {D_{\rm DW}}^{-1} \right\}_{\frac{N}{2},N} D_N .  \nonumber\\
\end{eqnarray}
From this and Eq.~(\ref{eq:breaking-Delta}), 
we infer that the second and third terms of the right-hand side of
Eq.~(\ref{relation-of-axial-vector-currents-N}) 
vanish in the limit of the infinite flavors 
and we finally obtain 
Eq.~(\ref{relation-of-axial-vector-currents}). 

\subsection{Axial anomaly}
The flavor singlet axial vector current of the domain-wall fermion with
the subtraction of the Pauli-Villars field satisfies the 
following axial Ward-Takahashi identity:
\begin{eqnarray}
\label{eq:axial-WT-identity-DW-PV}
&& \partial_\mu^\ast \left\langle \overline{A}_{\mu {\rm DW}}(x) \right\rangle 
\nonumber\\
&& = 
2 \left\langle \bar \psi^\prime_{\frac{N}{2}}(x) \, \gamma_5 \, 
\psi^\prime_{\frac{N}{2}}(x) \right\rangle 
+ 
2 \left\langle \bar \phi^\prime_{\frac{N}{2}}(x) \, \gamma_5 \, 
\phi^\prime_{\frac{N}{2}}(x) \right\rangle 
+ 2 \left\langle \bar Q(x) \, \gamma_5 \, 
Q(x) \right\rangle  , \nonumber\\
\end{eqnarray}
where $Q(x)$ and $\bar Q(x)$ stand for the $N$-th flavor 
of the Pauli-Villars field as defined in Eq.~(\ref{eq:N-th-PV-field}).
Note that in this case three chiral symmetry breaking terms
would contribute to axial anomaly. The first term is the contribution
of the domain-wall fermion and was evaluated in 
section~\ref{sec:axial-anomaly-q-and-psi}. The second and third
terms are the contributions of the Pauli-Villars field. 
In order to evaluate them, we need the propagators of the
$\frac{N}{2}$-th flavor and $N$-th flavor of the Pauli-Villars field.
The propagator of $Q(x)$ and $\bar Q(x)$ was evaluated in
section~\ref{sec:effective-action-of-q}.
\begin{equation}
\langle Q(x) \bar Q(y) \rangle 
= \frac{1}{a^4}\frac{1}{D_N^{\rm eff}+ \frac{1}{a_5}}
= \frac{a_5 }{a^4}\left( 1- a D_N \right) .
\end{equation}
The propagator of the 
$\frac{N}{2}$-th flavor of the Pauli-Villars field
can be evaluated using the similar method to evaluate the 
propagator of the 
$\frac{N}{2}$-th flavor of the domain-wall fermion,  
which was used in section~\ref{sec:axial-anomaly-q-and-psi}.
The calculation is described in 
appendix~\ref{sec:propagators-heavy-modes}.
The result can be written as 
\begin{equation}
\left\langle \phi^\prime_{\frac{N}{2}}(x) 
\bar \phi^\prime_{\frac{N}{2}}(y)\right\rangle
= -\gamma_5 \frac{a_5 }{a^4} \left( 1-aD_N\right) \gamma_5 .
\end{equation}
From these results, we see that 
the net contribution of the anomalous terms of 
the r.h.s. of Eq.~(\ref{eq:axial-WT-identity-DW-PV}) is same
as that of the domain-wall fermion (without the subtraction of the 
Pauli-Villars field) and is given by
\begin{equation}
\label{eq:axial-WT-identity-DW-PV-result}
\partial_\mu^\ast \left\langle \overline{A}_{\mu {\rm DW}}(x) \right\rangle 
= \frac{1}{a^4} \left(
- a{\rm tr} \gamma_5 R D_N (x,x) - {\rm tr} D_N^{-1}  \Delta_N (x,x)
\right) .
\end{equation}

This identity corresponds term by term 
to the axial Ward-Takahashi identity of the truncated overlap Dirac
fermion, which is derived under the chiral transformation 
Eq.~(\ref{eq:chiral-transformation-of-psi-truncated}):
\begin{equation}
\label{eq:axial-WT-identity-truncated-overlap-Dirac-fermion}
\partial_\mu^\ast \left\langle A_{\mu N}(x) \right\rangle 
= \frac{1}{a^4} \left(
- a{\rm tr} \gamma_5 R D_N (x,x) 
+ \left\langle \bar \psi(x) \Delta_N \psi(x) \right\rangle 
\right) .
\end{equation}
And it reduces in the limit of the infinite flavors to
the anomalous axial Ward-Takahashi identity of the overlap
Dirac fermion:
\begin{equation}
\label{eq:axial-WT-identity-overlap-Dirac-fermion}
\partial_\mu^\ast \left\langle \widetilde A_{\mu }(x) \right\rangle 
= \frac{1}{a^4} \left(
- a{\rm tr} \gamma_5 R \widetilde D (x,x) 
\right) .
\end{equation}

\subsection{Probe for topological charge at a finite flavor}
In view of the Eqs.~(\ref{eq:axial-WT-ideinty-DW-mqq}),
(\ref{eq:anomalous-term-q}) and (\ref{eq:anomalus-term-DW-and-index}),
it seems reasonable to probe the topological charge by the anomalous term 
of Eq.~(\ref{eq:axial-WT-identity-DW}), as in 
the numerical calculation by Argonne group \cite{lagae-sinclair}.
In this respect, 
we may  probe the topological charge rather directly
using the truncated overlap Dirac operator. 
\begin{equation}
Q_N= -\frac{1}{2} a {\rm Tr} \gamma_5 R D_N .
\end{equation}
The topological charge density,
$-\frac{1}{2} a {\rm tr} \gamma_5 R D_N(x,x)$ is assumed to be
a local functional of the gauge fields 
in the limit of the infinite flavors 
\cite{index-theorem-at-finite-lattice,
exact-chiral-symmetry,locality-of-overlap-D}.
This fact is reflected in the domain-wall fermion
in that $Q_N$ can be evaluated as
a contribution of the massive flavor to the anomalous term
of Eq.~(\ref{eq:axial-WT-identity-DW}). 
We point out that $Q_N$ can also be
expressed as follows:
\begin{eqnarray}
Q_N 
&=& - a^4 \sum_x m  \left\langle \bar q (x) \gamma_5 q(x)  \right\rangle 
\Big\vert_{ma=1}  \nonumber\\
&=& a^4\sum_x \left\langle \bar Q (x) \gamma_5 Q(x) \right\rangle .
\end{eqnarray}

\section{Discussion}

We have discussed the chiral property of the light fermion
of the domain-wall fermion by considering its low energy effective
action 
\begin{equation}
  S_N^{\rm eff} = a^4 \sum_x \bar q(x) \, D_N^{\rm eff} \, q(x),
\end{equation}
where the effective Dirac operator has a simple relation
to the truncated overlap Dirac operator as
\begin{eqnarray}
\frac{a}{a_5} { D_N^{\rm eff} }^{-1}+ a \delta(x,y) 
&=& 
{ D_N^{\rm \phantom{f}} }^{-1} .
\end{eqnarray}
We have argued that the chiral property of 
the light fermion field is understandable also 
from the point of view of the exact chiral symmetry 
based on the Ginsparg-Wilson relation, which holds true
for the overlap Dirac operator. 

As discussed in section \ref{sec:effective-action-of-q}, 
in the limit of the infinite flavors,
the effective action itself becomes chiral, but non-local. 
A subtlety of the effective action 
in this limit becomes clear when one attempts to calculate the axial
anomaly directly from the effective action: with such chiral and 
non-local action, it is not easy to identify the source of the axial anomaly. 
One possible way might be to consider the explicit breaking term 
at a finite flavor $N$:
\begin{equation}
a^4 \sum_x  \alpha(x) \, 
\left\langle \bar q(x) \left\{ \gamma_5, D_N^{\rm eff} \right\} q (x) 
\right\rangle \Big\vert_{ma} .
\end{equation}
(We have introduced a bare mass of the light fermion and 
$\alpha(x)$ is an infinitesimal local parameter.)
In view of Eq.~(\ref{eq:anomalous-term-q}) and the axial
Ward-Takahashi identity, 
as least for the global case 
this term actually reproduces the anomaly as 
the index of the overlap Dirac operator in the limit 
of the infinite flavors.
We leave this question for future study.

\section*{Acknowledgments}
We would like to thank A.~Yamada, T.~Onogi, 
S.~Aoki, T.~Izubuchi and Y.~Taniguchi for enlightening discussions.
Y.K. is also grateful to O.~Miyamura and A.~Nakamura 
for discussions.
Y.K. is supported in part by Grant-in-Aid for Scientific Research from
Ministry of Education, Science and Culture(\#10740116,\#10140214).

\appendix
\section{Evaluation of propagators of heavy fermions and bosons}
\label{sec:propagators-heavy-modes}

In this appendix, 
we evaluate the correlation function between the $N$-th flavor and 
the $\frac{N}{2}$-th flavor of the domain-wall fermion
\begin{equation}
\left\langle \psi^\prime_{\frac{N}{2}}(x) \, \bar q(y) \right\rangle ,
\end{equation}
by the method used in 
section~\ref{sec:effective-action-of-q-calculation}.
This correlation function is used in the evaluation of the vacuum 
expectation value of the axial current in 
section \ref{sec:chiral-property-q-and-psi}. 
We also evaluate the correlation function of 
the $\frac{N}{2}$-th flavor of the domain-wall fermion and
that of the Pauli-Villars field
\begin{equation}
\left\langle \psi^\prime_{\frac{N}{2}}(x) 
\bar \psi^\prime_{\frac{N}{2}}(y) 
\right\rangle ,
\end{equation}
\begin{equation}
\left\langle \phi^\prime_{\frac{N}{2}}(x) 
\bar \phi^\prime_{\frac{N}{2}}(y) 
\right\rangle .
\end{equation}
These are necessary in the evaluation of the axial anomaly 
in sections \ref{sec:axial-anomaly-q-and-psi} and
\ref{sec:Pauli-Villars-field}. We note that the correlation function 
for the Pauli-Villars field is obtained from the result 
for the domain-wall fermion by a simple replacement 
\begin{eqnarray}
\label{eq:replace-DW-PV}
\alpha_0
= \frac{1}{a_5}
\left( \begin{array}{cc} B & -C^\dagger  \\
                                   0 & 0 \end{array} \right)
&\Longrightarrow&
\alpha_1
= \frac{1}{a_5}
\left( \begin{array}{cc} B & -C^\dagger  \\
                                   0 & 1 \end{array} \right),  \\
\beta_0
= \frac{1}{a_5}
\left( \begin{array}{cc} 
0 & 0 \\
C & B 
\end{array} \right) 
&\Longrightarrow&
\beta_1
= \frac{1}{a_5}
\left( \begin{array}{cc} 
1 & 0 \\
C & B 
\end{array} \right) .
\end{eqnarray}

For this purpose, we introduce the source terms for 
the $\frac{N}{2}$-th flavor fermion in addition to 
those for the light $N$-th flavor,
\begin{equation}
 a^4 \sum_x \left\{ 
\bar J_{\frac{N}{2}}(x) \psi_{\frac{N}{2}} (x) 
+ \bar  \psi_{\frac{N}{2}}(x) J_{\frac{N}{2}}(x) \right\} .
\end{equation}
In the upper-triangle basis of section 
\ref{sec:effective-action-of-q}, 
these source terms may be expressed as follows:
\begin{eqnarray}
&&  a^4 \sum_x \left\{ \bar J_{\frac{N}{2}} (x)
\left[ 
 P_R \left(\begin{array}{cc} 0 & 1 \\ 1 & 0 \end{array}\right) 
 \psi''_{\frac{N}{2}}(x)
+P_L \left(\begin{array}{cc} 0 & 1 \\ 1 & 0 \end{array}\right) 
 \psi''_{\frac{N}{2}+1}(x) \right] 
\right.  \nonumber\\
&& \qquad\qquad \qquad\qquad \qquad\qquad \qquad \qquad
\left.
+ \bar \psi''_{\frac{N}{2}} (x) 
\left(\begin{array}{cc} 0 & 1 \\ 1 & 0 \end{array}\right)
J_{\frac{N}{2}}(x) \right\} . \nonumber\\
\end{eqnarray}

After the integration of the first $\frac{N}{2}-1$ flavors, 
the terms which contains the $\frac{N}{2}$-th flavor are found as follows:
\begin{eqnarray}
&&
\left[
\bar J(x)  P_L \left(\begin{array}{cc} 0 & 1 \\ 1 & 0 \end{array}\right) 
+ 
\bar \psi''_N (x) \beta_0 
\psi''_{\frac{N}{2}} (x)
\right]
\left( - \alpha^{-1} \beta \right)^{\frac{N}{2}-1} 
\psi''_{\frac{N}{2}}(x)
\nonumber\\
&& \qquad\qquad\qquad\quad
+ \bar \psi''_{\frac{N}{2}}(x) \alpha \psi''_{\frac{N}{2}}(x)
\nonumber\\
&& \qquad\qquad\qquad\qquad\qquad
+ \bar \psi''_{\frac{N}{2}}(x) \beta \psi''_{\frac{N}{2}+1}(x)
\nonumber\\
&& \qquad\qquad
+ \bar J_{\frac{N}{2}} (x)
 P_R \left(\begin{array}{cc} 0 & 1 \\ 1 & 0 \end{array}\right) 
 \psi''_{\frac{N}{2}}(x)
+ \bar \psi''_{\frac{N}{2}}(x) 
\left(\begin{array}{cc} 0 & 1 \\ 1 & 0 \end{array}\right)
J_{\frac{N}{2}}(x) . 
\nonumber\\
\end{eqnarray}
Integrating the $\frac{N}{2}$-th flavor, we obtain
\begin{eqnarray}
&&
\left\{ 
\left[
\bar J(x)  P_L \left(\begin{array}{cc} 0 & 1 \\ 1 & 0 \end{array}\right) 
+ 
\bar \psi''_N (x) \beta_0 
\right]
\left( - \alpha^{-1} \beta \right)^{\frac{N}{2}-1} 
\right. \nonumber\\
&& \left. \qquad\qquad\qquad\qquad\qquad
+ 
\bar J_{\frac{N}{2}} (x)
 P_R \left(\begin{array}{cc} 0 & 1 \\ 1 & 0 \end{array}\right) 
\right\} \, 
\left( - \alpha^{-1} \right) 
\left(\begin{array}{cc} 0 & 1 \\ 1 & 0 \end{array}\right) 
J_{\frac{N}{2}} (x) 
\nonumber\\
&&
+
\left\{ 
\left[
\bar J(x)  P_L \left(\begin{array}{cc} 0 & 1 \\ 1 & 0 \end{array}\right) 
+
\bar \psi''_N (x) \beta_0 
\right]
\left(-\alpha^{-1}\beta\right)^{\frac{N}{2}}
\right.
\nonumber\\
&& \qquad\qquad
\left.
+ 
\bar J_{\frac{N}{2}} (x)
\left[
 P_R \left(\begin{array}{cc} 0 & 1 \\ 1 & 0 \end{array}\right) 
\left( - \alpha^{-1} \beta \right) 
+
P_L \left(\begin{array}{cc} 0 & 1 \\ 1 & 0 \end{array}\right) 
\right]
\right\}
\psi''_{\frac{N}{2}+1} (x)  
\nonumber\\
&& \qquad\qquad\qquad\qquad
+ \bar \psi''_{\frac{N}{2}+1}(x) \, \alpha \, 
\psi''_{\frac{N}{2}+1}(x)
\nonumber\\
&& \qquad\qquad\qquad\qquad\qquad\qquad
+ \bar \psi''_{\frac{N}{2}+1}(x) \, \beta \, \psi''_{\frac{N}{2}+2}(x) .
\end{eqnarray}
Further integration up to the $N-1$-th flavor results in 
\begin{eqnarray}
&& 
\bar J(x) \left[
 P_L \left(\begin{array}{cc} 0 & 1 \\ 1 & 0 \end{array}\right) 
\left( - \alpha^{-1} \beta \right)^{N-1}
+P_R \left(\begin{array}{cc} 0 & 1 \\ 1 & 0 \end{array}\right) 
 \right] \psi''_N(x)
\nonumber\\
&& \qquad\qquad 
+ 
\bar \psi''_N(x) \, 
\left[ \alpha_0
      +\beta_0 \, 
\left( - \alpha^{-1} \beta \right)^{N-1} \right] \psi''_N(x) 
+ \bar \psi''_N (x) 
\left(\begin{array}{cc} 0 & 1 \\ 1 & 0 \end{array}\right)
J(x)
\nonumber\\
&&
+\bar J_{\frac{N}{2}} (x)
\left[
 P_R \left(\begin{array}{cc} 0 & 1 \\ 1 & 0 \end{array}\right) 
\left( - \alpha^{-1} \beta \right) 
+
P_L \left(\begin{array}{cc} 0 & 1 \\ 1 & 0 \end{array}\right) 
\right]
\left( - \alpha^{-1} \beta \right)^{\frac{N}{2}-1} 
\psi''_N(x)  
\nonumber\\
&& \qquad\qquad\qquad\qquad\qquad\quad
+ \bar \psi''_N (x) \beta_0 
\left(-\alpha^{-1}\beta\right)^{\frac{N}{2}-1}
\left( - \alpha^{-1} \right) 
\left(\begin{array}{cc} 0 & 1 \\ 1 & 0 \end{array}\right) 
J_{\frac{N}{2}} (x) 
\nonumber\\
&&
+ \left\{ 
\bar J(x)  P_L \left(\begin{array}{cc} 0 & 1 \\ 1 & 0 \end{array}\right) 
\left( - \alpha^{-1} \beta \right)^{\frac{N}{2}-1} 
+ 
\bar J_{\frac{N}{2}} (x)
 P_R \left(\begin{array}{cc} 0 & 1 \\ 1 & 0 \end{array}\right) 
\right\} \times \nonumber\\
&& \qquad \qquad \qquad \qquad \qquad \qquad \qquad 
\left( - \alpha^{-1} \right) 
\left(\begin{array}{cc} 0 & 1 \\ 1 & 0 \end{array}\right) 
J_{\frac{N}{2}} (x) .
\end{eqnarray}

Noting Eq.~(\ref{eq:alpha-beta-T});
\begin{eqnarray}
\label{eq:alpha-beta-T-appendix}
\alpha_0 \alpha^{-1} &=& 
\left(\begin{array}{cc} 0 & 1 \\ 1 & 0 \end{array}\right) 
P_L 
\left(\begin{array}{cc} 0 & 1 \\ 1 & 0 \end{array}\right),  \nonumber\\
\beta_0 \beta^{-1} &=& 
\left(\begin{array}{cc} 0 & 1 \\ 1 & 0 \end{array}\right) 
P_R 
\left(\begin{array}{cc} 0 & 1 \\ 1 & 0 \end{array}\right),  \\
\left(-\beta \, \alpha^{-1}\right) 
&=&
\left(\begin{array}{cc} 0 & 1 \\ 1 & 0 \end{array}\right) 
\gamma_5 \, e^{a_5 \widetilde H} \, \gamma_5 
\left(\begin{array}{cc} 0 & 1 \\ 1 & 0 \end{array}\right) ,
\end{eqnarray}
we can evaluate the factor 
of the first term as follows:
\begin{eqnarray}
&&
\left[
 P_R \left(\begin{array}{cc} 0 & 1 \\ 1 & 0 \end{array}\right) 
\left( - \alpha^{-1} \beta \right) 
+
P_L \left(\begin{array}{cc} 0 & 1 \\ 1 & 0 \end{array}\right) 
\right]
\left( - \alpha^{-1} \beta \right)^{\frac{N}{2}-1}  \alpha^{-1}
\nonumber\\
&& \qquad \qquad = 
a_5 e^{ \frac{N}{2} a_5 \widetilde H } 
(-\gamma_5)
\left(\begin{array}{cc} 0 & 1 \\ 1 & 0 \end{array}\right) .
\end{eqnarray}
Combining with Eq.~(\ref{eq:factor-1-2});
\begin{eqnarray}
&&
\left[
 P_L \left(\begin{array}{cc} 0 & 1 \\ 1 & 0 \end{array}\right) 
\left( - \alpha^{-1} \beta \right)^{N-1}
+P_R \left(\begin{array}{cc} 0 & 1 \\ 1 & 0 \end{array}\right) 
 \right] \alpha^{-1}
\nonumber\\
&& \qquad \qquad = - a_5
\left[ P_R + P_L \, e^{N a_5 \widetilde H } \right] (-\gamma_5)
\left(\begin{array}{cc} 0 & 1 \\ 1 & 0 \end{array}\right) ,  \\
\label{eq:factor-N-N-DW}
&& \left[ \alpha_0
      +\beta_0 \, 
\left( - \alpha^{-1} \beta \right)^{N-1} \right] \alpha^{-1}
\nonumber\\
&& \qquad \qquad 
= 
\left(\begin{array}{cc} 0 & 1 \\ 1 & 0 \end{array}\right) 
\left[ P_L + P_R \, e^{N a_5 \widetilde H } \right] (-\gamma_5)
\left(\begin{array}{cc} 0 & 1 \\ 1 & 0 \end{array}\right) ,
\end{eqnarray}
we obtain
\begin{eqnarray}
\left\langle \psi'_{\frac{N}{2}}(x) \, \bar q(y) \right\rangle 
&=& \frac{a_5}{a^4} \,
e^{ \frac{N}{2} a_5 \widetilde H } 
\frac{1}{P_L + P_R \, e^{N a_5 \widetilde H}} \nonumber\\
&=& \frac{a_5}{a^4}\,
\frac{1}{ \cosh \frac{N}{2} a_5 \widetilde H } \, \, 
\frac{1}{1+\gamma_5 \tanh \frac{N}{2} a_5 \widetilde H } .
\end{eqnarray}
As to the correlation function of the $\frac{N}{2}$-th flavor, we obtain
\begin{eqnarray}
\label{eq:correlation-function-half-flavor}
\left\langle \psi'_{\frac{N}{2}}(x) 
\bar \psi'_{\frac{N}{2}}(y) 
\right\rangle 
&=& 
\frac{a_5}{a^4} \left(
-P_R + 
e^{ \frac{N}{2} a_5 \widetilde H } 
\frac{1}{P_L + P_R \, e^{N a_5 \widetilde H}} 
P_R 
e^{ \frac{N}{2} a_5 \widetilde H }  \gamma_5  \right) \nonumber\\
&=& \frac{a_5}{a^4} \left(
-P_R + 
\frac{1}{ P_L \, e^{-\frac{N}{2} a_5 \widetilde H} 
         +P_R \, e^{\frac{N}{2} a_5 \widetilde H}} \,
P_R \, 
e^{ \frac{N}{2} a_5 \widetilde H }  \gamma_5  \right) \nonumber\\
&=& \frac{a_5}{a^4} \left(
-P_L - 
\frac{1}{ P_L \, e^{-\frac{N}{2} a_5 \widetilde H} 
         +P_R \, e^{\frac{N}{2} a_5 \widetilde H}} \,
P_L \, 
e^{ - \frac{N}{2} a_5 \widetilde H }  \gamma_5  \right) . \nonumber\\
\end{eqnarray}

By averaging the last two expressions of 
Eq.~(\ref{eq:correlation-function-half-flavor}), 
we have
\begin{eqnarray}
\left\langle \psi'_{\frac{N}{2}}(x) 
\bar \psi'_{\frac{N}{2}}(y) 
\right\rangle 
&=& \frac{a_5}{a^4} \left\{ -\frac{1}{2}
+ \frac{1}{2} \, \frac{1}{\cosh \frac{N}{2} a_5 \widetilde H }  \times
\right.
\nonumber\\
&& \qquad \qquad 
\, \,   \frac{1}{1+\gamma_5 \tanh \frac{N}{2} a_5 \widetilde H } 
\, \gamma_5 \, 
\left( 1+\gamma_5 \tanh \frac{N}{2} a_5 \widetilde H \right) \times
\nonumber\\
&& \qquad \qquad \qquad  
\left.
\, \, \cosh \frac{N}{2} a_5 \widetilde H 
\gamma_5 \right\} . 
\end{eqnarray}
We then remind the following identity, 
\begin{equation}
\gamma_5 D_N = - D_N \gamma_5 \left(1- a R D_N\right) + \Delta_N ,
\end{equation}
where
\begin{equation}
D_N = \frac{1}{2a} 
\left(1+\gamma_5 \tanh \frac{N}{2} a_5 \widetilde H \right) 
\end{equation}
\begin{equation}
\gamma_5 \left(1- a R D_N\right) = - \tanh \frac{N}{2} a_5 \widetilde H ,
\end{equation}
and
\begin{equation}
\Delta_N = \gamma_5 \frac{1}{a} 
\frac{1}{\cosh^2 \frac{N}{2} a_5 \widetilde H }.
\end{equation}
Then we finally obtain the following expression 
\begin{eqnarray}
\left\langle \psi^\prime_{\frac{N}{2}}(x) \, 
\bar \psi^\prime_{\frac{N}{2}}(y) 
\right\rangle 
&=& \frac{a_5}{a^4} \left(-1 + \frac{1}{2} \gamma_5 a R D_N \gamma_5 
\right. \nonumber\\
&& \qquad\qquad \left.
+ \frac{1}{2a}\frac{1}{\cosh \frac{N}{2} a_5 \widetilde H } 
\, \,   \frac{1}{ D_N } \, \gamma_5 \, \, 
\frac{1}{\cosh \frac{N}{2} a_5 \widetilde H } \, \gamma_5  \right) .
\nonumber\\
\end{eqnarray}

In the case of the Pauli-Villars field, 
through the replacement of Eq.~(\ref{eq:replace-DW-PV}), 
we obtain
\begin{eqnarray}
\alpha_1 \alpha^{-1} &=& 
\left(\begin{array}{cc} 0 & 1 \\ 1 & 0 \end{array}\right) 
\left(-\gamma_5\right)
\left(\begin{array}{cc} 0 & 1 \\ 1 & 0 \end{array}\right),  \nonumber\\
\beta_1 \beta^{-1} &=& 
\left(\begin{array}{cc} 0 & 1 \\ 1 & 0 \end{array}\right) 
\left(+\gamma_5\right)
\left(\begin{array}{cc} 0 & 1 \\ 1 & 0 \end{array}\right)
\end{eqnarray}
and
\begin{eqnarray}
&& \left[ \alpha_1
      +\beta_1 \, 
\left( - \alpha^{-1} \beta \right)^{N-1} \right] \alpha^{-1}
\nonumber\\
&& \qquad \qquad 
= 
\left(\begin{array}{cc} 0 & 1 \\ 1 & 0 \end{array}\right) 
\left[  1 + e^{N a_5 \widetilde H } \right] (-\gamma_5)
\left(\begin{array}{cc} 0 & 1 \\ 1 & 0 \end{array}\right) ,
\end{eqnarray}
instead of Eqs.~(\ref{eq:alpha-beta-T-appendix}) and
(\ref{eq:factor-N-N-DW}).
Then we can evaluate the correlation function as
\begin{eqnarray}
\left\langle \phi'_{\frac{N}{2}}(x) 
\bar \phi'_{\frac{N}{2}}(y) 
\right\rangle 
&=& 
\frac{a_5}{a^4} \left(
-P_R + 
e^{ \frac{N}{2} a_5 \widetilde H } 
\frac{1}{1 + \, e^{N a_5 \widetilde H}} 
e^{ \frac{N}{2} a_5 \widetilde H }  \gamma_5  \right) \nonumber\\
&=&
\frac{a_5}{a^4} \left(
 -\frac{1}{2}+ \frac{1}{2} \tanh \frac{N}{2} a_5 \widetilde H
     \, \gamma_5 \right) \nonumber\\
&=&
- \gamma_5 
\frac{a_5}{a^4} \left(1-a D_N \right) \gamma_5 .
\end{eqnarray}

\end{document}